\documentclass[12pt,pre,aps,showpacs,preprint]{revtex4}

\usepackage{graphics}
\usepackage{graphicx}
\usepackage{amsfonts}
\usepackage{textcomp}
\usepackage{amssymb}
\usepackage{amsmath}

\begin{document}

\title{Geometrical Ambiguity of Pair Statistics. II. Heterogeneous Media}

\author{Yang Jiao}


\affiliation{Department of Mechanical and Aerospace Engineering,
Princeton University, Princeton New Jersey 08544, USA}

\author{Frank H. Stillinger}


\affiliation{Department of Chemistry, Princeton University,
Princeton New Jersey 08544, USA}

\author{Salvatore Torquato}

\email{torquato@electron.princeton.edu}

\affiliation{Department of Chemistry, Princeton University,
Princeton New Jersey 08544, USA}

\affiliation{Department of Physics, Princeton University,
Princeton New Jersey 08544, USA}

\affiliation{Princeton Institute for the Science and Technology of
Materials, Princeton University, Princeton New Jersey 08544, USA}

\affiliation{Program in Applied and Computational Mathematics,
Princeton University, Princeton New Jersey 08544, USA}

\affiliation{Princeton Center for Theoretical Science, Princeton
University, Princeton New Jersey 08544, USA}

\date{\today}


\begin{abstract}

In the first part of this series of two papers [Y. Jiao, F. H. Stillinger, and
S. Torquato, Phys. Rev. E {\bf 81}, 011105 (2010)], we considered the
geometrical ambiguity of pair statistics associated with point
configurations. Here we focus on the analogous problem for
heterogeneous media (materials). Heterogeneous media are
ubiquitous in a host of contexts, including composites and
granular media, biological tissues, ecological patterns and
astrophysical structures. The complex structures of heterogeneous
media are usually characterized via statistical descriptors, such
as the $n$-point correlation function $S_n$. An intricate inverse
problem of practical importance is to what extent a medium can be
reconstructed from the two-point correlation function $S_2$ of a
target medium. Recently, general claims of the uniqueness of
reconstructions using $S_2$ have been made based on numerical
studies, which implies that $S_2$ suffices to uniquely determine
the structure of a medium within certain numerical accuracy. In
this paper, we provide a systematic approach to characterize the
geometrical ambiguity of $S_2$ for both continuous two-phase
heterogeneous media and their digitized representations in a
mathematically precise way. In particular, we derive the exact
conditions for the case where two distinct media possess identical
$S_2$, i.e., they form a degenerate pair. The degeneracy
conditions are given in terms of integral and algebraic equations
for continuous media and their digitized representations,
respectively. By examining these equations and constructing
their rigorous solutions for specific examples, we conclusively
show that in general $S_2$ is indeed not sufficient information to
uniquely determine the structure of the medium, which is
consistent with the results of our recent study on heterogeneous
media reconstruction [Jiao, Stillinger and Torquato, Proc. Nat.
Acad. Sci. {\bf 106}, 17634 (2009)]. The analytical examples
include complex patterns composed of building blocks bearing the
letter ``T'' and the word ``WATER'' as well as degenerate stacking
variants of the densest sphere packing in three dimensions (Barlow
films). Several numerical examples of degeneracy (e.g.,
reconstructions of polycrystal microstructures, laser-speckle
patterns and sphere packings) are also given, which are virtually
exact solutions of the degeneracy equations. The uniqueness issue
of multiphase media reconstructions and additional structural
information required to characterize heterogeneous media are
discussed, including two-point quantities that contain topological
connectedness information about the phases.


\end{abstract}

\maketitle

\section{Introduction}


Two-phase heterogeneous media (textures) abound in nature and
synthetic situations. Examples include manufactured heterogeneous materials
(e.g., composites, porous media and colloids)
\cite{torquato,Sa03,Zo06,Me07}, geologic media (e.g., rock formations)
\cite{torquato,Sa03,sandstone}, cellular
materials \cite{Gi99}, ecological structures (e.g., tree patterns in
forests) \cite{ecology}, cosmological structures (e.g., galaxy
distributions) \cite{Pe93,Ga05}, and biological media (e.g.,
animal and plant tissue) \cite{Kh08}. In general, the complex
microstructures of random media can only be characterized via
certain statistical descriptors, such as an infinite set of
$n$-point correlation functions $S_n$ ($n=1, 2, \ldots$)
\cite{torquato}. It is well known that the effective physical
properties of heterogeneous media, such as the conductivity
\cite{cond}, elastic moduli \cite{elast}, fluid permeability
\cite{fluid}, trapping constant \cite{trap} and electromagnetic wave
characteristics \cite{Re08}, can be
expressed in terms of weighted functionals of $S_n$. In
particular, $S_n({\bf x}_1, {\bf x}_2, \ldots, {\bf x}_n)$ gives
the probability of finding $n$ points positioned at ${\bf x}_1,
{\bf x}_2, \ldots, {\bf x}_n$ all in the phase of interest \cite{torquato}.
For statistically homogeneous media which are the focus of this
paper, $S_n$ is translationally invariant and hence depends only
on the relative displacements of the positions with respect to
some arbitrarily chosen origin, say ${\bf x}_1$, i.e., $S_n({\bf
x}_1,{\bf x}_2,\dots,{\bf x}_n) = S_n({\bf x}_{12},{\bf
x}_{13},\dots,{\bf x}_{1n})$ with ${\bf x}_{ij} = {\bf x}_j-{\bf
x}_i$. In such cases, the one-point correlation function $S_1$ is
just equal to the volume fraction of the phase of interest. If the
medium is also statistically isotropic, the two-point correlation
function $S_2$ is a radial function, i.e., $S_2({\bf x}_{12}) =
S_2(|{\bf x}_{12}|)$, providing information about the distribution
of pair-separation distances.

\begin{figure}[bthp]
\begin{center}
$\begin{array}{c@{\hspace{1.5cm}}c}\\
\includegraphics[width=4.5cm,keepaspectratio]{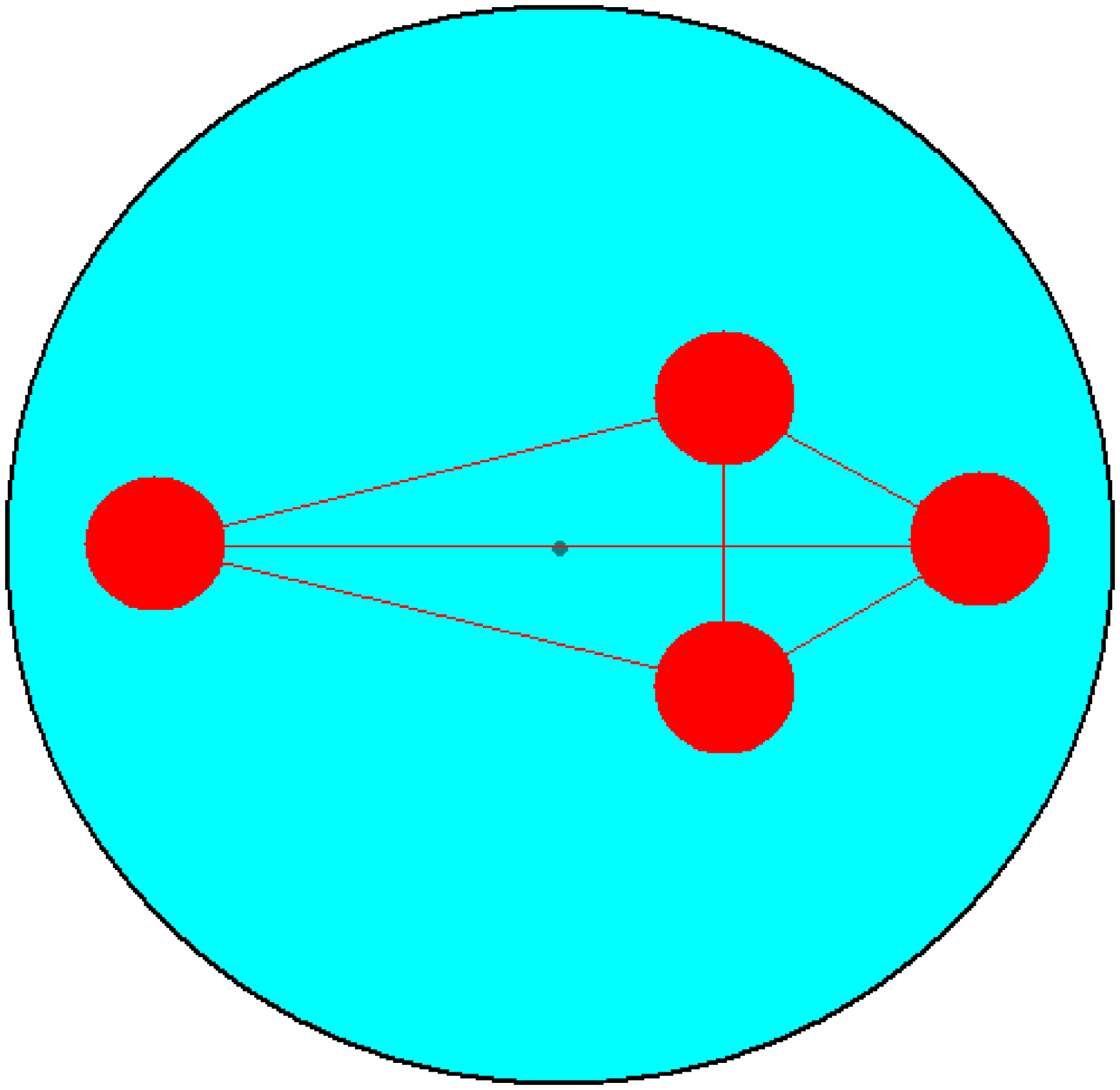} &
\includegraphics[width=4.5cm,keepaspectratio]{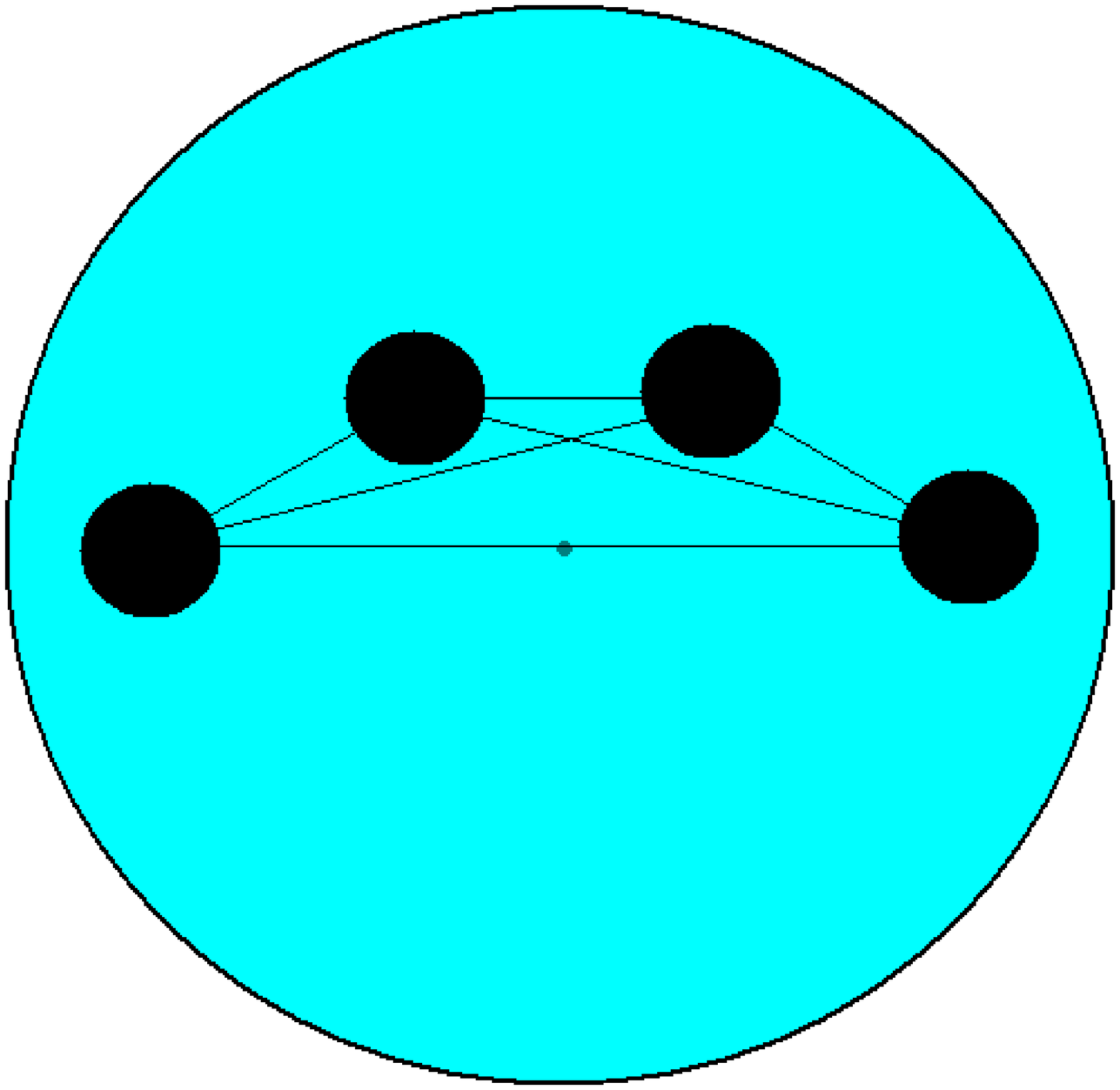} \\
\mbox{\bf (a)} & \mbox{\bf (b)}
\end{array}$
\end{center}
\caption{(color online). A degenerate pair of two-phase continuous
media based on the ``kite-trapezoid" example given in
Ref.~\cite{Ji09a}. The longest distance in the ``kite'' and
``trapezoid'' is symmetrically placed on the large circle
diameter, as adapted from Ref.~\cite{Ji09a}. Note that this
example can be interpreted as two-dimensional (disks) or
three-dimensional (spheres).} \label{fig1}
\end{figure}

Although the $S_n$ can be represented
analytically for certain models \cite{torquato} and bounded
for general media \cite{To83}, it is usually not
possible to compute  all of the $S_n$ in the infinite set.
Thus, it is desirable to understand the extent
to which one can  characterize the structure and properties of
heterogeneous media using lower-order correlation functions, such
as the two-point correlation function $S_2$. A powerful means to study this
problem and related questions  is to employ inverse techniques \cite{To09a}
whereby one attempts to reconstruct (or construct) realizations of heterogeneous media that
match limited structural information of those media in the form of
lower-order correlation functions, obtained either experimentally
or from theoretical considerations \cite{Ye98a, Ye98b}. In
particular, the reconstruction of digitized representations of
heterogeneous materials from a prescribed two-point correlation
function $S_2$ has been receiving considerable attention
\cite{Cu99, Utz02, ApplyA, ApplyB, ApplyC, ApplyD, ApplyE, Ji07, Ji08, Ka08}. An effective
reconstruction procedure enables one to generate accurate
renditions of the medium at will and subsequent analysis can be
performed on the reconstruction to obtain desired macroscopic
properties of the medium non-destructively.


It has been well established \cite{Ye98a, Ji07, Ba04, Yel93,pnas}
that $S_2$ is generally devoid of crucial information to uniquely
determine the structure of the medium. In particular, in
Ref.~\cite{pnas} we employed  inverse ``reconstruction''
techniques to probe the information content of the widest class of
different types of two-point functions. This set of functions
includes the standard two-point correlation function $S_2$,
surface-void $F_{sv}$ and surface-surface $F_{ss}$ correlation
functions \cite{surface}, lineal-path function $L$
\cite{lineal-path}, chord-length probability density function $p$
\cite{To93}, the pore-size function $F$ \cite{torquato}, and the
two-point cluster function $C_2$ \cite{cluster}. By numerically
reconstructing two-phase heterogeneous media from these
correlation functions, we unambiguously showed that $S_2$ does not
suffice for a unique reconstruction and that incorporating $C_2$,
which is sensitive to topological connectedness information, can
lead to a much more accurate rendition of the target medium. It is
worth noting that in most numerical reconstructions, $S_2$ of the
reconstructed medium only matches that of the target medium within
certain numerical accuracy \cite{accuracy}. However, it appears
that the idea that $S_2$ is generally devoid of crucial structural
information still has not been widely appreciated and claims of
uniqueness of reconstructions using $S_2$ have been made based on
numerical studies \cite{Utz02, Ka08}. Thus, our work has
implications for the fundamental problem of determining the
necessary conditions that realizable two-point functions must
possess \cite{To99}.

In the first part of this series of two papers \cite{Ji09a} (henceforth
referred to as Part I), we have shown via concrete examples the
existence of distinct point configurations possessing identical
sets of pair-separation distances. By appropriately decorating the
point configurations, distinct two-phase media with identical
$S_2$ can be constructed (see Fig.~\ref{fig1}). In general, there
are many ways to decorate a point configuration, which enables one
to obtain a wide spectrum of distinct heterogeneous media.
In this paper,  we focus on characterizing the structural ambiguity of pair
statistics as embodied in  the two-point correlation function $S_2$
associated with individual realizations of statistically
homogeneous two-phase heterogeneous media in a \textit{mathematically
precise} way, which complements  previous numerical studies.
In other words, we provide a mathematical formulation for
heterogeneous-media reconstructions and rigorously show that
distinct media exist that possess exactly identical two-point correlation functions.
This objective is quite different  from the approximate  matching of  $S_2$
that one encounters in numerical simulations, which are also discussed.

The reconstruction of random textures also has a wide spectrum of applications
in vision research \cite{Yel92, Yel93, Ch00}. It is important to distinguish a
popular two-point statistics used in vision research, called the ``dipole histogram'',
from the two-point correlation function of statistically homogeneous media.
The dipole histogram is a particular statistics of the vector pair displacements
in a \textit{finite} texture \cite{Ch00}. In such a case, the
distinguishable vector pair displacement with the largest
length can always be identified together with the positions of the two points
that give rise to the displacement. Similarly, one can then identify the
second largest displacement and the positions of points possessing this displacement, and so on.
Finally, the texture can be completely reconstructed from the dipole histogram.
However, we emphasize that this is possible because the information contained in the dipole histogram
of a finite texture is not averaged out as that contained in the two-point correlation
function of a statistically homogeneous medium. In particular, due to the requirement of
translational invariance, periodic boundary conditions are always imposed
for realizations of statistically homogeneous media. Thus, only the probabilities
of the occurrence of displacements are given by $S_2$, obtained by averaging
over all available positions in the medium, instead of the indication of
the actual occurrence of particular displacements at certain positions provided by the
dipole histogram. It is such additional information contained in the dipole histogram
that enables a unique reconstruction. We will see in the following sections that
the position average (i.e., volume average) necessarily leads to the structural
ambiguity of $S_2$.


\begin{figure}[bthp]
\begin{center}
$\begin{array}{c}\\
\includegraphics[width=12.0cm,keepaspectratio]{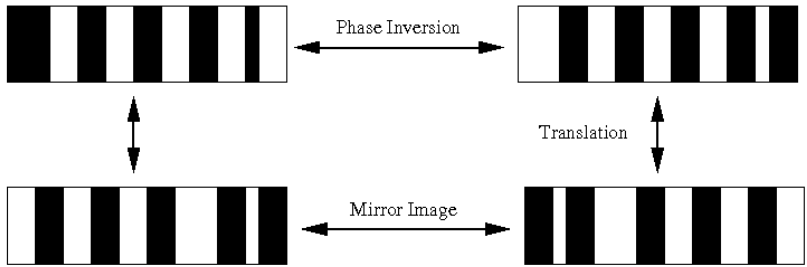} \\
\end{array}$
\end{center}
\caption{Illustrations of one-dimensional two-phase random media
that form trivial degenerate pairs that we do not count. Note that
the arrows indicate that the transformations (e.g., mirror
reflection, translation, phase inversion and these combinations)
can occur in either the clockwise or counter-clockwise
directions.} \label{fig2}
\end{figure}

To provide a quantitative characterization of the structural
ambiguity of $S_2$, we first give the relevant definitions here.
Two heterogeneous media are identical if they possess identical
sets of $n$-point correlation functions $S_n$ for $n=1, 2,
\ldots$. Two heterogeneous media are $S_k$-distinct if they
process distinct $n$-point correlation functions for all $n \ge
k$. A heterogeneous medium is $l$-fold degenerate if there exist
additional $(l-1)$ media such that all the $l$ media are mutually
$S_3$-distinct but possess the same two-point correlation function
$S_2$. For statistically homogeneous media, the above definition
of structural degeneracy rules out the possibility that two
degenerate media are trivially connected by a translation,
rotation, mirror reflection or any of these combinations (see
Fig.~\ref{fig2}). This is consistent with our definition of
degeneracy for point configuration in the first part of this
series two papers \cite{Ji09a}. Moreover, we consider that two
media do not form a degenerate pair if they possess
phase-inversion symmetry at phase volume fractions 0.5, i.e., the
morphology of one phase is statistically identical to that of the
other phase when the two phases are interchanged \cite{torquato}.


As pointed out in Part I, given pair statistics in the form of
$S_2$ that are associated with degenerate media, it is impossible
even in principle to uniquely reconstruct a medium from such
$S_2$. Furthermore, it is important to determine under what
conditions the media possess degenerate counterparts. In this
paper, we provide a systematic approach to characterize the
structural degeneracies associated with the two-point correlation
function for both continuous two-phase heterogeneous media and
their digitized representations. In particular, we derive the
exact conditions for the existence of degeneracy in terms of
integral equations for continuous media and algebraic equations
for the digitized representations. We show by examining the
derived equations that for statistically homogeneous and isotropic
media, structural degeneracies generally exist. This explains the
long observed non-uniqueness of reconstructions using radial
averaged $S_2$ in many studies \cite{Ye98a, Ye98b,Cu99, Ji07,
Ji08}. We also provide a variety of concrete examples of
degenerate two-phase media, including both analytical
constructions and numerical simulations, which respectively are
exact and approximate solutions of the degeneracy equations.
These examples include analytically constructed patterns composed
of building blocks bearing the letter ``T'' and the word
``WATER'', degenerate Barlow films (i.e., stacking variants of the
densest sphere packing in three dimensions) as well as numerical
reconstructions of polycrystal microstructures, laser-speckle
patterns and sphere packings. Moreover, we discuss the degeneracy
of multiphase media and additional structural information that can
be incorporated to significantly reduce structural degeneracy of
heterogeneous media \cite{pnas}.


The rest of the paper is organized as follows: In Sec.~II, we
introduce the mathematical model of two-phase heterogeneous media
and derive the conditions of degeneracy. In Sec.~III, we provide
the examples of degeneracies for both continuous media and their
digitized representations. In Sec.~IV, we discuss the degeneracy
of multiphase media and additional structural information that
would be used to reduce degeneracy. In Sec.~V, we make concluding
remarks.

\section{Conditions of Degeneracy}

In this section, we derive the exact mathematical conditions under which a statistically
homogeneous medium possesses degenerate counterparts. For a continuous medium,
these conditions take the form of a set of integral equations of the indicator
function (defined below) of the phase of interest. For the digitized representations
of the medium, which have been extensively investigated in various numerical reconstruction
studies, the integral equations reduce to algebraic equations.

\subsection{Continuous Media}

Consider a statistically homogeneous medium $M$ occupying the region ${\cal V}$ in the
$d$-dimensional Euclidean space $\mathbb{R}^d$ ($d=1,2,3$) which is partitioned into two
disjoint phases \cite{torquato}: phase 1, a region ${\cal V}_1$ of volume fraction $\phi_1$ and phase 2,
a region ${\cal V}_2$ of volume fraction $\phi_2$. It's obvious that ${\cal V}_1 \cup {\cal V}_2 = {\cal V}$
and ${\cal V}_1 \cap {\cal V}_2 = {\bf 0}$. The indicator function ${\cal I}^{(i)}({\bf x})$ of phase
$i$ is given by

\begin{equation}
\label{eq00201}{\cal I}^{(i)}\left( {\bf x} \right) = \left\{
{{\begin{array}{*{20}c}
 {1 \quad\quad {\rm {\bf x}} \in {\cal V}_i}, \\
 {0 \quad\quad {\rm {\bf x}} \in \bar {{\cal V}_i}},
\end{array} }} \right.
\end{equation}

\noindent for $i = 1,2$ with ${\cal V}_i \cup \bar {{\cal V}_i} = {\cal V}$ and

\begin{equation}
\label{eq00202}{\cal I}^{(1)}({\bf x}) + {\cal I}^{(2)}({\bf x}) = 1.
\end{equation}

\noindent The $n$-point correlation function $S^{(i)}_n$ for phase $i$ is defined as follows:

\begin{equation}
\label{eq00203} S^{(i)}_n ({\bf x}_1,{\bf x}_2,...,{\bf x}_n,) =
\left\langle{{\cal I}^{(i)}({\bf x}_1){\cal I}^{(i)}({\bf x}_2)...{\cal I}^{(i)}({\bf x}_n)}\right\rangle,
\end{equation}

\noindent where the angular brackets ``$\left\langle{...}\right\rangle$'' denote ensemble
averaging over independent realizations of the medium. The two-point correlation function $S^{(i)}_2$
for phase $i$ is defined by

\begin{equation}
\label{eq00204} S^{(i)}_2({\bf x}_1,{\bf x}_2) =
\left\langle{{\cal I}^{(i)}({\bf x}_1){\cal I}^{(i)}({\bf x}_2)}\right\rangle.
\end{equation}

\noindent As pointed out in Sec.~I, for a statistically homogeneous medium, $S_2^{(i)}$ is a function
of the relative displacements of point pairs, i.e.,

\begin{equation}
\label{eq00205} S^{(i)}_2({\bf x}_1,{\bf x}_2) = S^{(i)}_2({\bf x}_2-{\bf x}_1) = S^{(i)}_2({\bf r}),
\end{equation}

\noindent where ${\bf r} = {\bf x}_2-{\bf x}_1$. In the infinite volume limit, if the medium is also ergodic
the ensemble average is equivalent to the volume average, i.e.,

\begin{equation}
\label{eq00206} S^{(i)}_2({\bf r}) = \lim_{V\rightarrow\infty} \frac{1}{V}\int_V {{\cal I}^{(i)}\left(
{\rm {\bf x}} \right){\cal I}^{(i)}\left( {{\rm {\bf x}} + {\rm {\bf r}}}
\right)d{\rm {\bf x}}}
\end{equation}

\noindent If the medium is also statistically isotropic, $S_2^{(i)}$ is a radial function, depending
on the separation distances of point pairs only, i.e.,

 \begin{equation}
\label{eq00207} S^{(i)}_2({\bf x}_1,{\bf x}_2) = S^{(i)}_2(|{\bf r}|) = S^{(i)}_2(r).
\end{equation}

\noindent Readers are referred to Ref.~\cite{torquato} for a detailed discussion of $S_2^{(i)}$
and other higher order $S_n^{(i)}$. Henceforth, we will drop the superscript $i$ in $S_2^{(i)}$
for simplicity. Without further elaboration, $S_2$ is always the two-point correlation function
of the phase of interest.

Now consider a change in the geometry of the region of the phase of interest, say phase 1,
that keeps the volume fraction $\phi_1$ invariant. The medium after the change is denoted by $M'$
and recall that the original medium is denoted by $M$. The indicator function ${\cal I}'$ of phase 1
in $M'$ is then given by

\begin{equation}
\label{eq202} {\cal I}'\left( {\rm {\bf x}} \right) = \left\{
{{\begin{array}{*{20}c}
 {1 \quad\quad {\rm {\bf x}} \in {\cal V}'_1}, \\
 {0 \quad\quad {\rm {\bf x}} \in \bar {{\cal V}'_1}},
\end{array} }} \right.
\end{equation}

\noindent where ${\cal V}'_1$ is the region where phase 1 is found in $M'$.
We define the change of indicator function as:

\begin{equation}
\label{eq203} \delta {\cal I}\left( {\rm {\bf x}} \right) = {\cal I}'\left( {\rm
{\bf x}} \right) - {\cal I}\left( {\rm {\bf x}} \right),
\end{equation}

\noindent where ${\cal I}$ is the indicator function of phase 1 in $M$ given by Eq.~(\ref{eq00201})
with $i=1$. It is obviously that $\delta {\cal I}$ can only take the value of 0, -1 and 1.
And the requirement that $M$ and $M'$ form a degenerate pair with the same two-point
correlation function as well as the same phase volume fractions will lead to constraining
equations on $\delta {\cal I}$, which are the degeneracy conditions we are seeking.

For simplicity, we introduce a short notation for the integral in Eq.~(\ref{eq00206}), i.e.,

\begin{equation}
\label{eq207} S_2 = \frac{1}{V}{\cal I} \otimes {\cal I}.
\end{equation}

\noindent The requirement that $S_2$ is invariant under the change $\delta I$ leads to

\begin{equation}
\label{eq01201} S'_2 = S_2 = \frac{1}{V}{\cal I}'\otimes {\cal I}' = \frac{1}{V} {\cal I}\otimes {\cal I},
\end{equation}

\noindent Substituting (\ref{eq203}) into (\ref{eq01201}) yields

\begin{equation}
\label{eq208} \frac{1}{V}\left( {{\cal I} + \delta {\cal I}} \right) \otimes
\left( {{\cal I} + \delta {\cal I}} \right) = \frac{1}{V}{\cal I} \otimes {\cal I}.
\end{equation}

\noindent From Eq.~(\ref{eq208}) we can obtain that

\begin{equation}
\label{eq209} \left( {{\cal I} + \frac{1}{2}\delta {\cal I}} \right) \otimes
\delta {\cal I} + \delta {\cal I} \otimes \left( {{\cal I} + \frac{1}{2}\delta {\cal I}}
\right) = 0.
\end{equation}

\noindent Note that given the indicator function ${\cal I}$ of the original medium $M$,
Eq.~(\ref{eq209}) should be satisfied by $\delta {\cal I}$ for all ${\bf r}$ within
the range of interest. In other words, (\ref{eq209}) specifies a continuous
spectrum of equations, one for each $\delta {\cal I}$ at a particular ${\bf r}$. Another
requirement for degeneracy is that the phase volume fractions are conserved, which
leads to

\begin{equation}
\label{eq204} \frac{1}{V}\int_V {{\cal I}'\left( {\rm {\bf x}}
\right)d{\rm {\bf x}}} = \frac{1}{V}\int_V {{\cal I}\left( {\rm {\bf x}}
\right)d{\rm {\bf x}}}.
\end{equation}

\noindent Substituting (\ref{eq203}) into (\ref{eq204}), we can obtain that

\begin{equation}
\label{eq205} \int_V {\delta {\cal I}\left( {\rm {\bf x}} \right)d{\rm {\bf x}}} = 0.
\end{equation}

\noindent If the medium $M$ is also statistically isotropic, $S_2$ depends only on pair-separation distances, i.e.,

\begin{equation}
\label{eq01202}S_2(r) = \frac{1}{\Omega}\int_{\Theta}S_2({\bf r})d\Theta
\end{equation}

\noindent where the integral over $d$-dimensional solid angle $\Theta$ is to average over all
directions of ${\bf r}$ with the same length $r$ and $\Omega$ is the total solid angle.
Thus from Eq.~(\ref{eq209}), we obtain that

\begin{equation}
\label{eq01203}\displaystyle{\int_{\Theta}\left[{ \left( {{\cal I} + \frac{1}{2}\delta {\cal I}} \right) \otimes
\delta {\cal I} + \delta {\cal I} \otimes \left( {{\cal I} + \frac{1}{2}\delta {\cal I}}
\right)} \right]d{\Theta} = 0}
\end{equation}

\noindent Note that (\ref{eq01203}) should be satisfied by $\delta {\cal I}$ for all ${r}$
within the range of interest.

Thus, it is clear that for the statistically homogeneous medium $M$, its structural
degeneracy $M'$ exists only if Eqs.~(\ref{eq209}) and (\ref{eq205}) possess
nontrivial solutions; and the degeneracy exists only if Eqs.~(\ref{eq01203}) and
(\ref{eq205}) possess nontrivial solutions if $M$ is also statistically isotropic.
In principle, from the solutions of Eqs.~(\ref{eq209}) [or (\ref{eq01203})] and (\ref{eq205}) together
with the indicator function ${\cal I}$ of the original medium, one can obtain the indicator function
${\cal I}'$ of the degenerate medium by applying (\ref{eq203}). Henceforth, we will call Eq.~(\ref{eq205})
the \textit{feasibility} condition, since it requires that $\delta I$ must be feasible in
the sense that the phase volume fractions are conserved. Similarly, we will call Eqs.~(\ref{eq209})
and (\ref{eq01203}) the \textit{invariance} condition, since they give a set of $\delta {\cal I}$ that
leave $S_2$ invariant under the change.

\subsection{Digitized Media}
For a digitized medium $M$ in $\mathbb{R}^d$, the indicator function takes the form of a finite $d$-dimensional
array ${\bf I} = [I_{x_1 \ldots x_d}]$ of linear size $N$, where $I_{x_1 \ldots x_d} = 0$ or $1$ indicating the
phase of the pixel at $(x_1 \ldots x_d)$ and $x_i = 1, \ldots, N$ for $i=1,\ldots, d$. Consider a perturbation of the geometry
of phase 1 in the digitized medium $M$ that preserves the phase volume fractions (e.g., by exchanging pixels of
different phases). The perturbed medium with the indicator function ${\bf I}' = [I'_{x_1 \ldots x_d}]$ is denoted
by $M'$. The change of the indicator function is simply

\begin{equation}
\delta{\bf I} = {\bf I}'-{\bf I} = [I'_{x_1 \ldots x_d} - I_{x_1 \ldots x_d}] = [\delta I_{x_1 \ldots x_d}],
\label{eq01}
\end{equation}

\noindent which is also a $d$-dimensional matrix with linear size $N$. The entries $\delta I_{x_1 \ldots x_d}$
can only take the values $-1, 0$ and $1$.

For statistically homogeneous digitized media, periodic boundary conditions are imposed by the requirement
of translational invariance. Thus, the indices of the indicator functions are modulated by $N$, i.e.,
if $(x_i + r_i) \ge N$ the index should take the value $(x_i +r_i -N)$ instead, where $x_i$ is the original
index and $r_i$ is the translation displacement along $x_i$ direction. On the other hand,
if $(x_i + r_i) < 0$ the index should take the value $(x_i +r_i +N)$ instead. The integral equations of
the feasibility condition (\ref{eq205}) reduce to algebraic equations, i.e.,

\begin{equation}
\displaystyle{\sum \limits_{x_1,\ldots, x_d} \delta I_{x_1 \ldots x_d} = 0,}
\label{eq02}
\end{equation}

\noindent where the sum is running through $x_i = 1, \ldots, N$ for $i=1, \ldots, d$. The equations for
the invariance condition reduce to

\begin{equation}
\displaystyle{\sum\limits_{(x_1,\ldots, x_d) \in H_1} \delta I_{x_1\ldots x_d} +
\sum\limits_{(x_1,\ldots, x_d)\in H_2} \delta I_{(x_1+r_1)\ldots(x_d+r_d)} +
\sum\limits_{x_1, \ldots, x_d} \delta I_{x_1\ldots x_d} \delta I_{(x_1+r_1)\ldots(x_d+r_d)} = 0},
\label{eq03}
\end{equation}

\noindent where $r_i$ ($i=1, \ldots, d$) are integers satisfying $|r_i| \le N/2$ and the
sets of pixel positions $H_1$ and $H_2$ are given by

\begin{equation}
\label{eq04}
\begin{array}{c}
\displaystyle{H_1 = \left\{{(x_1,\ldots, x_d)~|~I_{(x_1+r_1)\ldots(x_d+r_d)} = 1}\right\}, }\\\\
\displaystyle{H_2 = \left\{{(x_1,\ldots, x_d)~|~I_{x_1 \ldots x_d} = 1}\right\}}.
\end{array}
\end{equation}

If the medium is also statistically isotropic, following Eq.~(\ref{eq01203}) the
invariance conditions can be obtained by averaging Eq.~(\ref{eq03}) over equivalent directions, i.e.,

\begin{equation}
\label{eq05}
\displaystyle{\sum\limits_{(r_1,\ldots, r_d)\in\Omega}\left({\sum\limits_{(x_1,\ldots, x_d)\in
H_1} \delta I_{x_1\ldots x_d} + \sum\limits_{(x_1,\ldots, x_d) \in H_2} \delta
I_{(x_1+r_1)\ldots(x_d+r_d)} + \sum\limits_{x_1,\ldots, x_d} \delta I_{x_1\ldots x_d} \delta
I_{(x_1+r_1)\ldots(x_d+r_d)}}\right) = 0}.
\end{equation}

\noindent where the sets of displacements with the same length $\Omega$ and
the sets of pixel positions $H_1$ and $H_2$ are given by

\begin{equation}
\label{eq06}
\begin{array}{c}
\Omega = \left \{{(r_1, \ldots,r_d)~| ~r_1^2 +\cdots+ r_d^2 = r^2, r \in \mathbb{Z}, r \le N/2}\right\}, \\\\
H_1 = \left\{{(r_1, \ldots, r_d)~| ~I_{(x_1+r_1)\ldots(x_d+r_d)} = 1}\right\},\\\\
H_2 = \left\{{(r_1, \ldots, r_d)~| ~I_{x_1 \ldots x_d} = 1}\right\}.\\
\end{array}
\end{equation}

Thus, similar to their continuous counterparts, two statistically homogeneous digitized media $M$ and
$M'$ form a degenerate pair if the feasibility condition Eq.~(\ref{eq02}) and the invariance condition
Eq.~(\ref{eq03}) hold for some non-trivial $\delta{\bf I}$. If the media are also statistically isotropic,
the degeneracy exists if Eqs.~(\ref{eq02}) and (\ref{eq05}) possess non-trivial solutions.
Once $\delta{\bf I}$ is obtained, the degenerate medium $M'$ can be constructed by applying Eq.~(\ref{eq01}).



Note that in Eq.~(\ref{eq05}), because the number of equations (i.e., the total number of integers that are smaller than $N^2/4$) is
in general much smaller than the number of unknowns (i.e., the number of the entries $I_{x_1 \ldots x_d}$ which is
equal to $N^d$), the algebraic equations (\ref{eq05}) possess multiple solutions. For such isotropic media,
it is highly probable to obtain degeneracies of the original medium in a reconstruction. This explains
the long observed non-uniqueness issue in the reconstruction of heterogeneous media using radial $S_2$
\cite{Ye98a, Ye98b, Ji07, Ji08}. The situation
is more subtle for general statistically homogeneous media, for which the degeneracy condition Eq.~(\ref{eq03})
contains the same number of equations and unknowns. Thus, the chances for finding a non-trivial solution of
$\delta{\bf I}$ is much smaller but one still cannot rule out the possibility of degeneracy. It is clear that
the solutions also depend on the original medium, i.e., the values of $I_{x_1 \ldots x_d}$. In the following
section, we will show the existence of degeneracies for both continuous and digitized media via concrete examples,
which are solutions of the above degeneracy equations.

\section{Examples of Degeneracy}

In this section, we will provide concrete examples of structural
degeneracy. These examples include both exact and approximate
solutions of the equations for degeneracy derived in the previous
section. For continuous and simple digitized media, analytical
constructions are given. For more complicated digitized media,
numerical simulations are employed to find degeneracies. For a
clear illustration, we will mainly focus on two-dimensional
examples here. However, the general construction methods can be
also applied to find degenerate media in any space dimension.

\subsection{Analytical Constructions}

\subsubsection{Vector-Argument $S_2({\bf r})$}

\begin{figure}[bthp]
\begin{center}
$\begin{array}{c@{\hspace{1.5cm}}c}\\
\includegraphics[width=3.5cm,keepaspectratio]{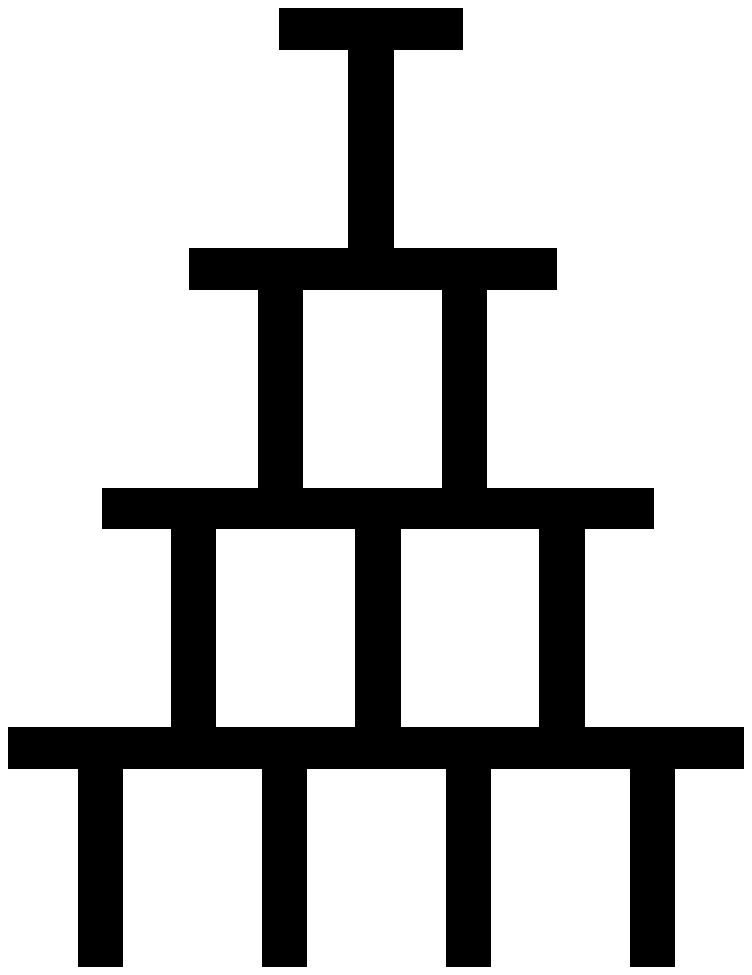} &
\includegraphics[width=3.5cm,keepaspectratio]{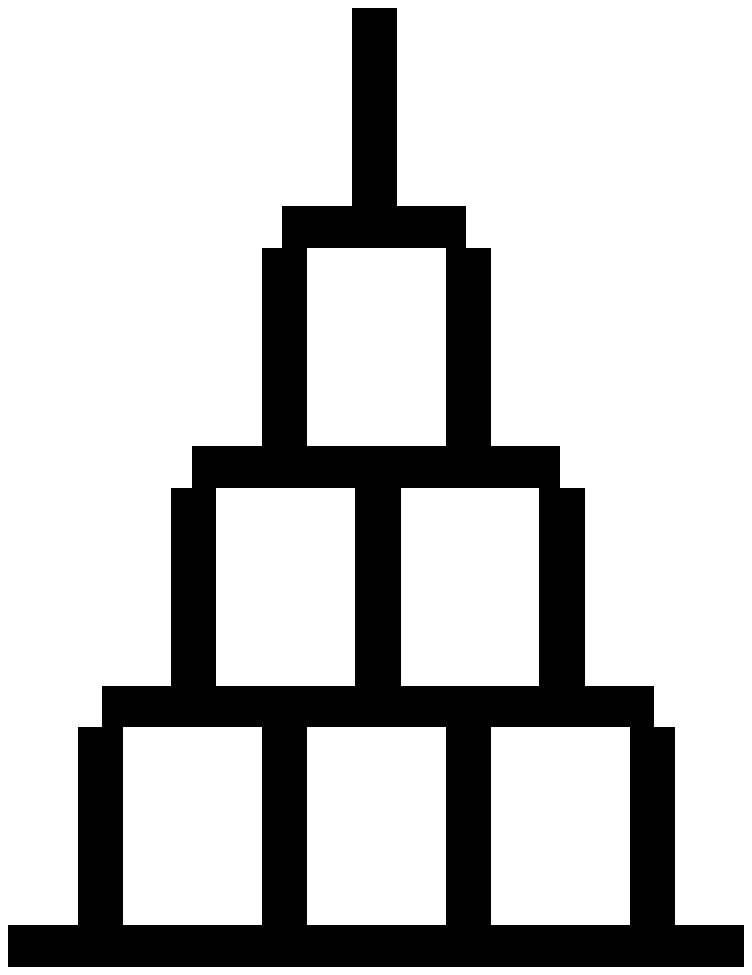} \\
\mbox{\bf (a)} & \mbox{\bf (b)}
\end{array}$
\end{center}
\caption{A simple degeneracy example composed of replications of letter ``T''.} \label{fig3}
\end{figure}

As discussed in Sec.~II.A, the two-point correlation function $S_2({\bf r})$ contains
information on the relative displacements (vector distances) between any two points in the
phase of interest. Consider a two-phase medium, in which one of the phases is composed of
replications of certain substructures. If the substructures do not possess central inversion
symmetry, they can be rotated in a way such that all the relative displacements of point
pairs within and among the substructures remain the same yet the overall structure
(the replication of the rotated substructures) is different from the original one.
A simple example originally proposed in Ref.~\cite{Yel92} is shown in Fig.~\ref{fig3}. The substructure,
i.e., the letter ``T'', has been arranged into a ``pyramid''. By rotating each ``T'' about
an axis perpendicular to the plane by 180 degrees, a distinct structure can be produced.
Since the relative orientation of the position of any pair of letters ``T'' has not changed, the
relative displacements of the point pairs, one from each of different ``T'' also remain the same.
It is obvious that the relative displacements of point pairs within any single ``T'' are
not affected. Thus, the two distinct structures possess identical statistics of relative
displacements of point pairs, i.e., identical two-point correlation functions $S_2({\bf r})$,
so they form a degenerate pair.

\begin{figure}[bthp]
\begin{center}
$\begin{array}{c}\\
\includegraphics[width=9.5cm,keepaspectratio]{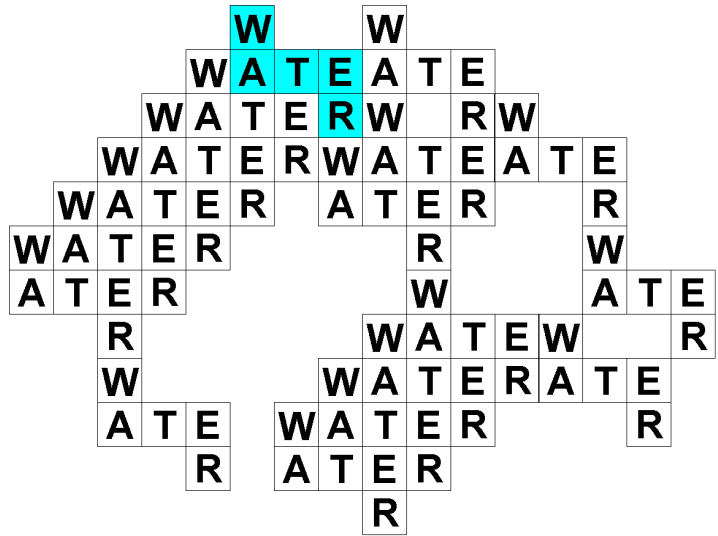} \\
\mbox{\bf (a)}\\\\
\includegraphics[width=9.5cm,keepaspectratio]{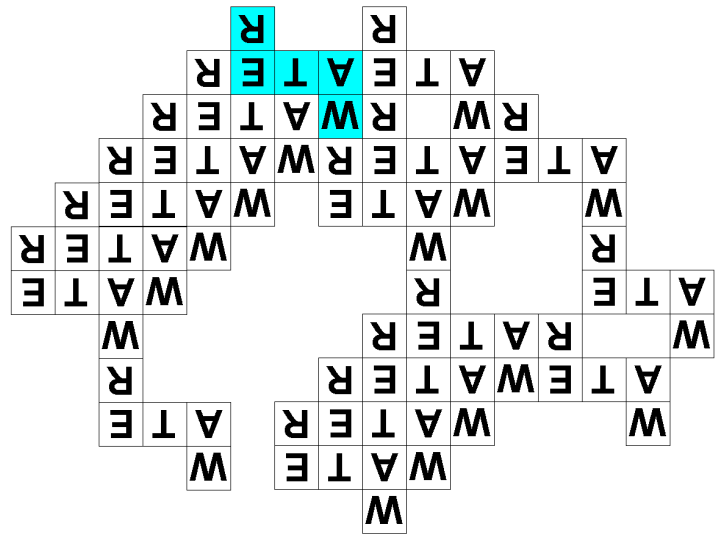} \\
\mbox{\bf (b)}
\end{array}$
\end{center}
\caption{A more sophisticated degeneracy example composed of
replications of the letters ``WATER''. One ``S''-shaped auxiliary
box is shown in blue (or light gray in the print version).}
\label{fig4}
\end{figure}

Fig.~\ref{fig4} shows a more sophisticated example based on the same
construction rule. The substructure now is the letters ``WATER'' arranged in
an ``S''-shaped auxiliary box (shown in blue),
which is then replicated in a complicated way. Note that though the auxiliary box possesses
central inversion symmetry, the substructure composed of the letters does not. By rotating
each substructure about the inversion symmetry axis of the corresponding auxiliary box,
a degenerate structure can be constructed. Note these examples are exact
solutions of the Eqs.~(\ref{eq205}) and (\ref{eq209}).

\subsubsection{Radially Averaged $S_2(r)$}

\begin{figure}[bthp]
\begin{center}
$\begin{array}{c@{\hspace{1.5cm}}c}\\
\includegraphics[width=4.5cm,keepaspectratio]{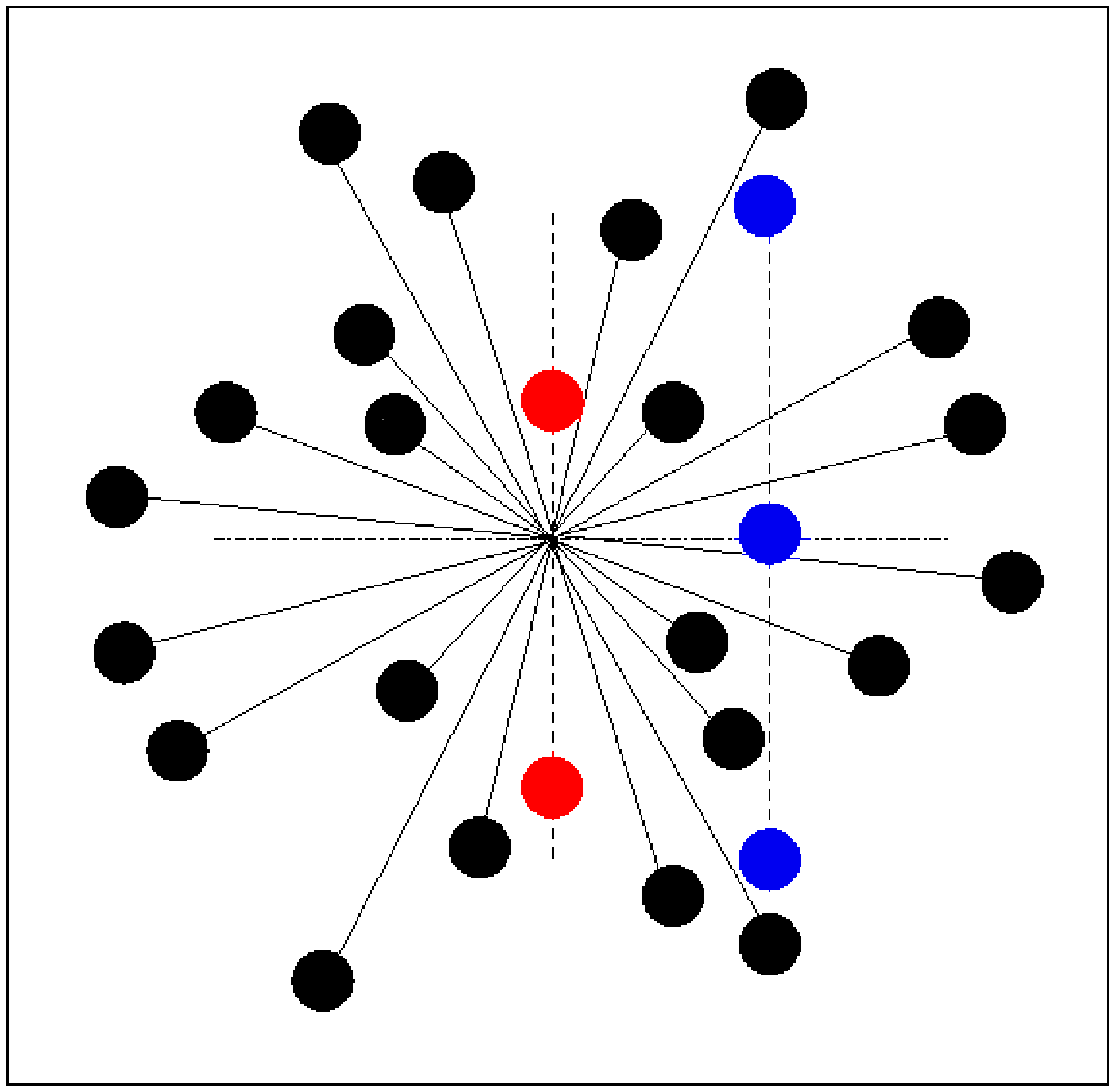} &
\includegraphics[width=4.5cm,keepaspectratio]{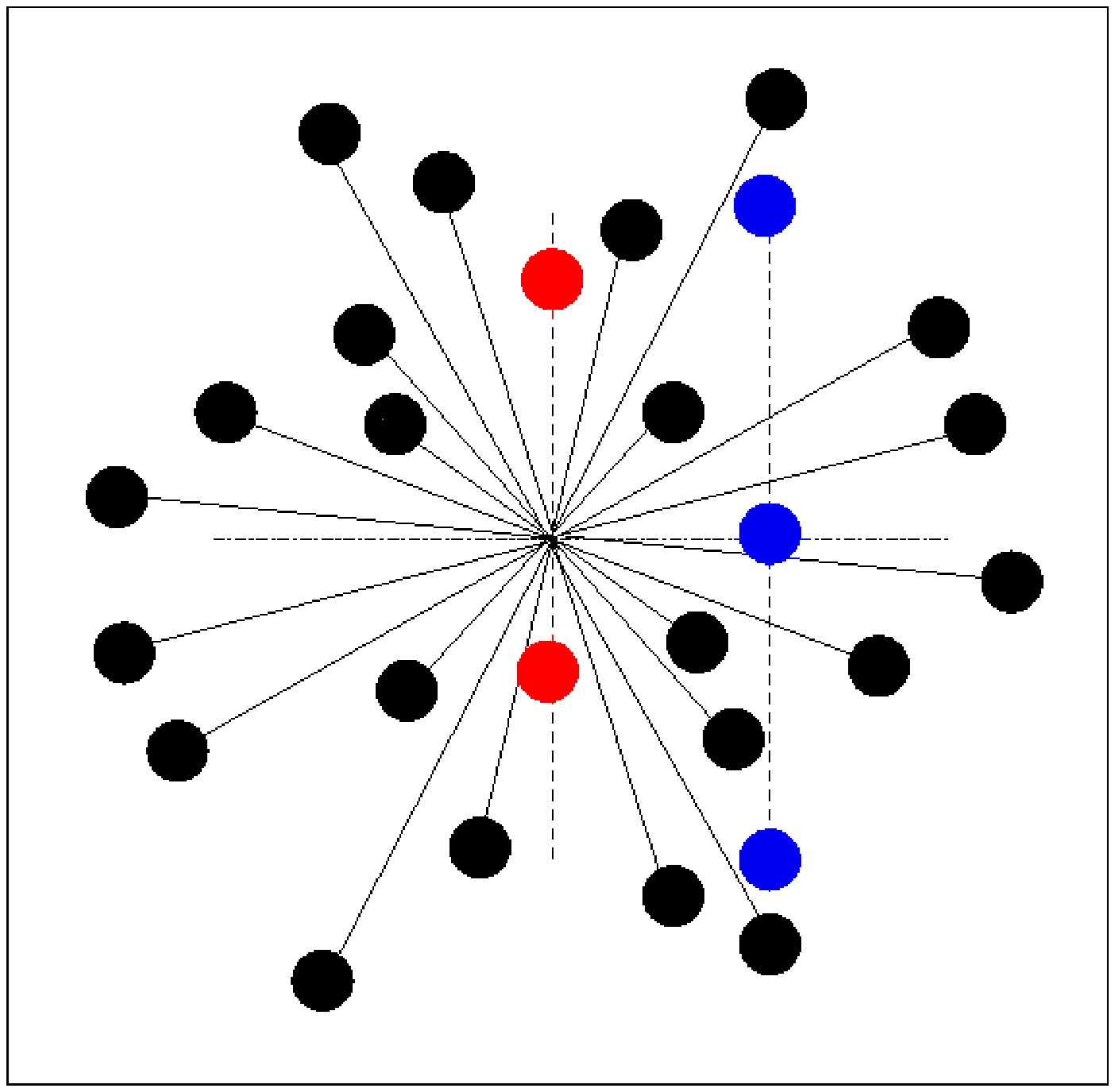} \\
\mbox{\bf (a)} & \mbox{\bf (b)}
\end{array}$
\end{center}
\caption{(color online). A degenerate pair of circular-disk
packings in two dimensions.} \label{fig5}
\end{figure}

Radially averaged $S_2(r)$ only contains information on separation distances
between the point pairs in the phase of interest. In the best case, the complete
set of pair-separation distances can be inferred from $S_2(r)$. Such distance
sets have been studied in detail in Part I of this series of two papers, where we
have shown that a variety of classes of degeneracies can be identified that
are compatible with the given distance set of a point configuration. In particular,
we pointed out in Part I that by decorating the degenerate point configurations properly,
degenerate two-phase media can be constructed. One such example is to decorate
the 30 degenerate tetrahedra discussed in Part I, e.g., one can place the centroids
of congruent spheres at the vertices of the tetrahedra and choose the ``sphere'' phase
as the phase of interest and the space exterior to spheres as ``void'' phase. Then
the 30 heterogeneous media associated these tetrahedra are degenerate.

Another example constructed by decorating a point configuration is shown
in Fig.~\ref{fig5}, where the two circular-disk packings form a degenerate pair. The packings
contain three types of particles: A large number of the circular disks (shown in black)
form a cluster with central symmetry. Three particles (shown in blue) arranged on a
straight line form a substructure possessing mirror reflection symmetry, i.e., the two most separated particles
are located symmetrically about the one in the middle. The remaining two (shown in red) are placed
along a line parallel to that of the three blue particles and going through the
center of the symmetric cluster, such that the set of relative distances between any
two points in the ``particle'' phase in the two packings are identical
while the two resulting structures are distinct. The constructed degenerate
pair is an exact solution of the Eqs.~(\ref{eq205}) and (\ref{eq01203}).

\begin{figure}[bthp]
\begin{center}
$\begin{array}{c@{\hspace{1.5cm}}c}\\
\includegraphics[width=4.5cm,keepaspectratio]{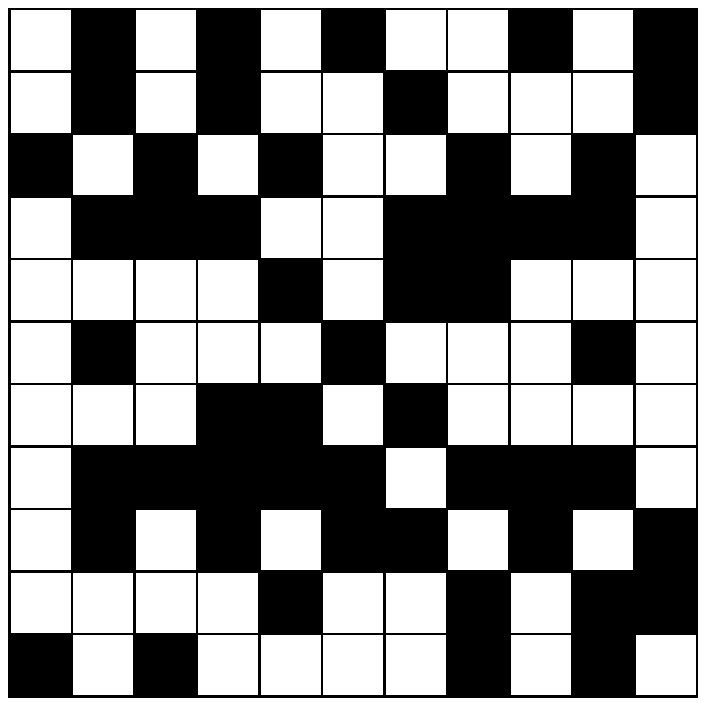} &
\includegraphics[width=4.5cm,keepaspectratio]{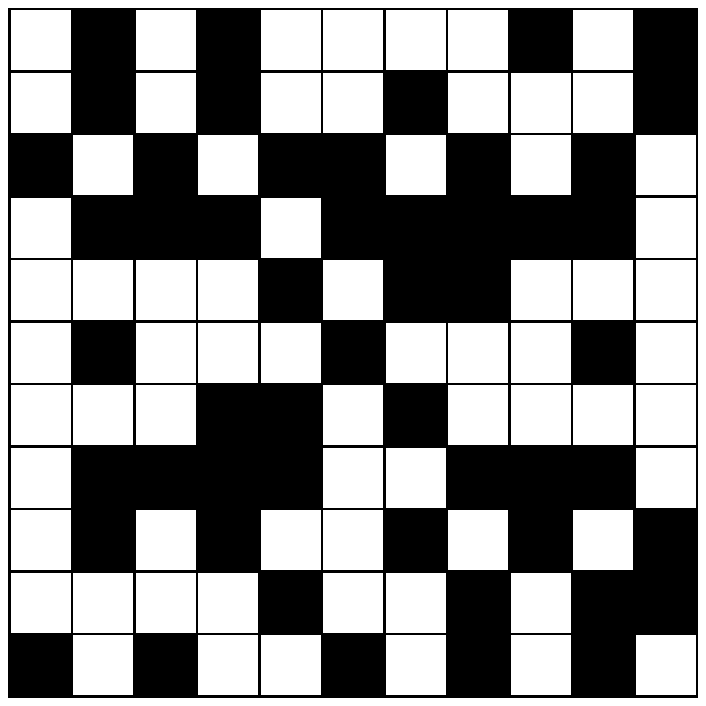} \\
\mbox{\bf (a)} & \mbox{\bf (b)}
\end{array}$
\end{center}
\caption{A degenerate pair of digitized media based on the same construction
rules for the degenerate circular-disk packings.} \label{fig6}
\end{figure}

Note that the aforementioned degeneracy construction rules are also applicable in the case
of digitized media. For example, applying the construction rule for the degenerate
circular-disk packings in Fig.~\ref{fig5} to a two-dimensional square-lattice of pixels results in the degenerate
digitized media shown in Fig.~\ref{fig6}. The only difference is an additional requirement that for
digitized media the values of the distances must be square roots of integers
and the symmetry groups of the structures are discrete, which reduces the total number of possible degeneracies.

\subsubsection{Degenerate Barlow Films}

Another class of interesting degeneracy examples associated with radially averaged $S_2$ involves the
Barlow packings, i.e., random stackings of infinite triangular-lattice layers of
spheres. In a Barlow packing, for a particular layer (i.e., layer ``A'') there are only
two possible positions for the next layer, which we denote by the usual code as ``B'' and ``C'', respectively.
Thus, a Barlow packing can be represented by a sequence of ``A'', ``B'' and ``C'', e.g.,
``$\ldots$ABCBCA$\ldots$''.
A Barlow film is a Barlow packing consisting of a finite number of layers.
The distribution of pair distances in such a geometry depends not only on
how many layers are present, but also upon whether any chosen pair of layers
is the same species (``AA'', ``BB'', or ``CC''), or different species
(``AB'', ``AC'', or ``BC'').  This motivates counting "same" versus "different" species
for each separation distance between layer pairs. The distance distribution (i.e., $S_2$) is
completely determined by the number of same species at all separation distances between
layer pairs.

For an $n$-layer Barlow film, there are nominally $2^n$ possible
configurations. However, not all of these $2^n$ configurations are
distinct. A basic property of Barlow films is that the distance
distribution is invariant under a ``renaming'' operation. For
example, the sequences ``ABCABC'', ``BCABCA'' and ``CABCAB''
possess the same structure. In light of this ``renaming''
invariance, we can require that the first two layers for all
Barlow films to be ``AB$\ldots$'' without loss of generality.
Thus, the number of distinct configurations associated with an
$n$-layer Barlow film that possess the same vertical distances
between identical layers (i.e., the same distance distributions)
is reduced to $2^{(n-2)}$. As mentioned in Sec.~I, reflection
(mirror image) symmetry should also be excluded for degenerate
pairs.

\begin{table}
\centering \caption{The number of the same species (``AA'', ``BB''
or ``CC'') at different layer-separation distances for the 6-layer
degenerate Barlow pair ``ABABAC'' and ``ABABCB''. $d_L$ is the
layer separation distance.}
\begin{tabular}{c@{\hspace{0.45cm}}c@{\hspace{0.45cm}}c}
\hline\hline
$d_L$ &  ``ABABAC'' & ``ABABCB''  \\
\hline
1    &  0   &  0  \\
2    &  3   &  3  \\
3    &  0   &  0  \\
4    &  1   &  1  \\
5    &  0   &  0   \\
\hline\hline
\end{tabular}
\label{tab1}
\end{table}

We then set out to identify degenerate Barlow film configurations up to $n=15$ layers by explicitly
generating all Barlow films and compare their layer separation distance distributions.
For $n<6$, there are no degenerate configurations.  An interesting case occurs
at $n=6$, with an inequivalent pair of Barlow films that have the same distance distribution,
and thus the same $S_2$ associated with the ``sphere'' phase.
They can be represented by ``ABABAC'' and ``ABABCB''. As shown in Table~\ref{tab1}, the two
configurations possess the same number of same layer species at all layer-separation distances. In
addition, they are not related by ``renaming'' and/or reflection operations.

\begin{table}
\centering \caption{The number of degenerate Barlow film configurations for $n \in [6,~15]$.
$n$ is the number of layers in the Barlow film. $N_{pair}$ is the number of degenerate pairs
and $N_{triplet}$ is the number of degenerate triplets.}
\begin{tabular}{c@{\hspace{0.45cm}}c@{\hspace{0.45cm}}c}
\hline\hline
 Number of Layers & $N_{pair}$ & $N_{triplet}$ \\
\hline
$n$=6    &   1   &  0  \\
$n$=7    &   1   &  0  \\
$n$=8    &   5   &  0  \\
$n$=9    &   4   &  0  \\
$n$=10   &   16   &  1  \\
$n$=11   &   13   &  0  \\
$n$=12   &   54   &  6  \\
$n$=13   &   35   &  2  \\
$n$=14   &   157   &  13  \\
$n$=15   &   123   &  5  \\
\hline\hline
\end{tabular}
\label{tab2}
\end{table}

The numbers of degenerate configurations for $n \in [6,~15]$ are given in Table~\ref{tab2}.
We note that the first degenerate triplet occurs at $n = 10$, which includes the configurations
``ABABCBACAC'', ``ABABCBCBAC'' and ``ABACBCBCAC''. We don't find any four-fold degenerate
Barlow films for the $n$  values we examined. It can be seen from the table that
except for the trivial cases with $n=6$ and $7$ the numbers of degeneracies possess the following trends:
i) for both the even-layer and the odd-layer films, the number and complexity of
degeneracies increase monotonically as the number of layers increases; (ii) the number of degeneracies
associated with an even layer is larger than that of the next odd layer. The latter trend is
probably due to the odd layers having relatively fewer degrees of freedom to exclude the
``renaming'' and reflection symmetry.

It is also worth noting that if a honeycomb crystal packing of spheres are used
for the layers instead of triangular-lattice packings, one can construct
degenerate low-density jammed sphere packings \cite{ToJam07}.
In particular, a honeycomb crystal layer can be obtained by removing one third
of the spheres from the triangular-lattice layer. These removed spheres are
also arranged on a triangular lattice, but with lattice vectors
that are twice in magnitude of those of the original triangular lattice. To obtain
honeycomb crystal films from two degenerate Barlow films, exactly the same
pair distances are removed associated with the removing of the spheres. Thus,
the two constructed honeycomb crystal films also form a degenerate pair.
So far, we have not found any degeneracy in the periodic Barlow films.

\subsection{Numerical Examples}

In many practical applications,
one encounters the issue of reconstructing the structures of
materials from radially-averaged two-point function $S_2(r)$, which might be the only available
information at hand and it is always associated with an inevitable
uncertainty. Therefore, it is also important to explore the
structural degeneracy of $S_2(r)$ subject to small uncertainties.
In this section, we employ the Yeong-Torquato heterogeneous material reconstruction
procedure \cite{Ye98a, Ye98b} to obtain numerical-degeneracy examples of
statistically homogeneous and isotropic media, which are
approximate solutions of the invariance equations (\ref{eq05})
derived in Sec.~II.B. {\bf In particular, the ``energy'' or \textit{mean squared error} $E$
for the reconstruction problem which has been defined previously \cite{Ye98a, Ye98b, Ji07, Ji08, pnas}
is given by
\begin{equation}
\label{eq_err}
E = \sum_r [\hat{S_2}(r) - S_2(r)]^2,
\end{equation}
where $\hat{S_2}$ and $S_2$ are the target and the reconstructed correlation functions,
respectively. In the Yeong-Torquato procedure, a stochastic optimization
technique is used to make $E$ in principle zero. However, in practice $E$ is
always a very small number, which corresponds to the situation
where the uncertainty associated with $S_2$ is very small.}


\begin{figure}[bthp]
\begin{center}
$\begin{array}{c@{\hspace{1.5cm}}c}\\
\includegraphics[width=4.5cm,keepaspectratio]{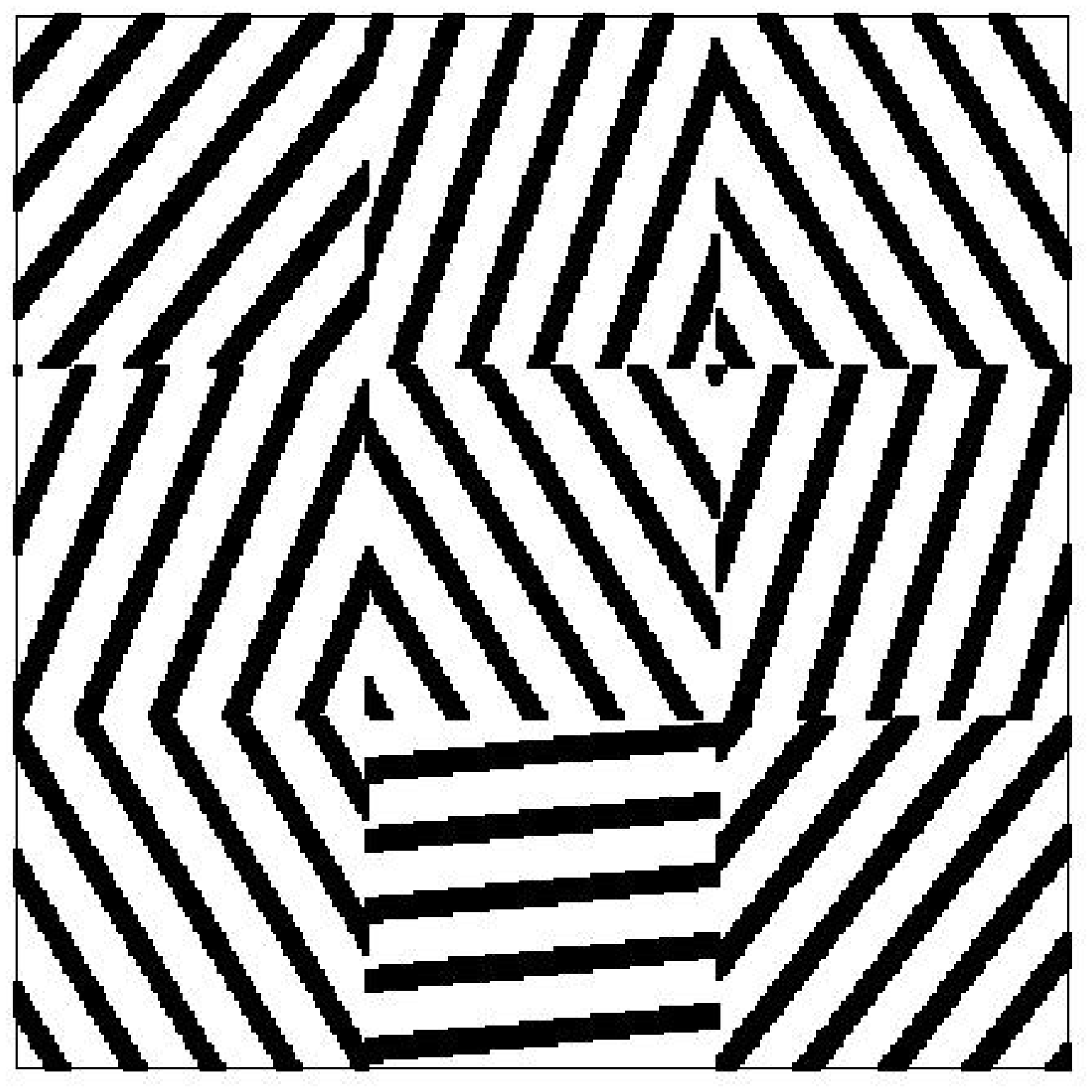} &
\includegraphics[width=4.5cm,keepaspectratio]{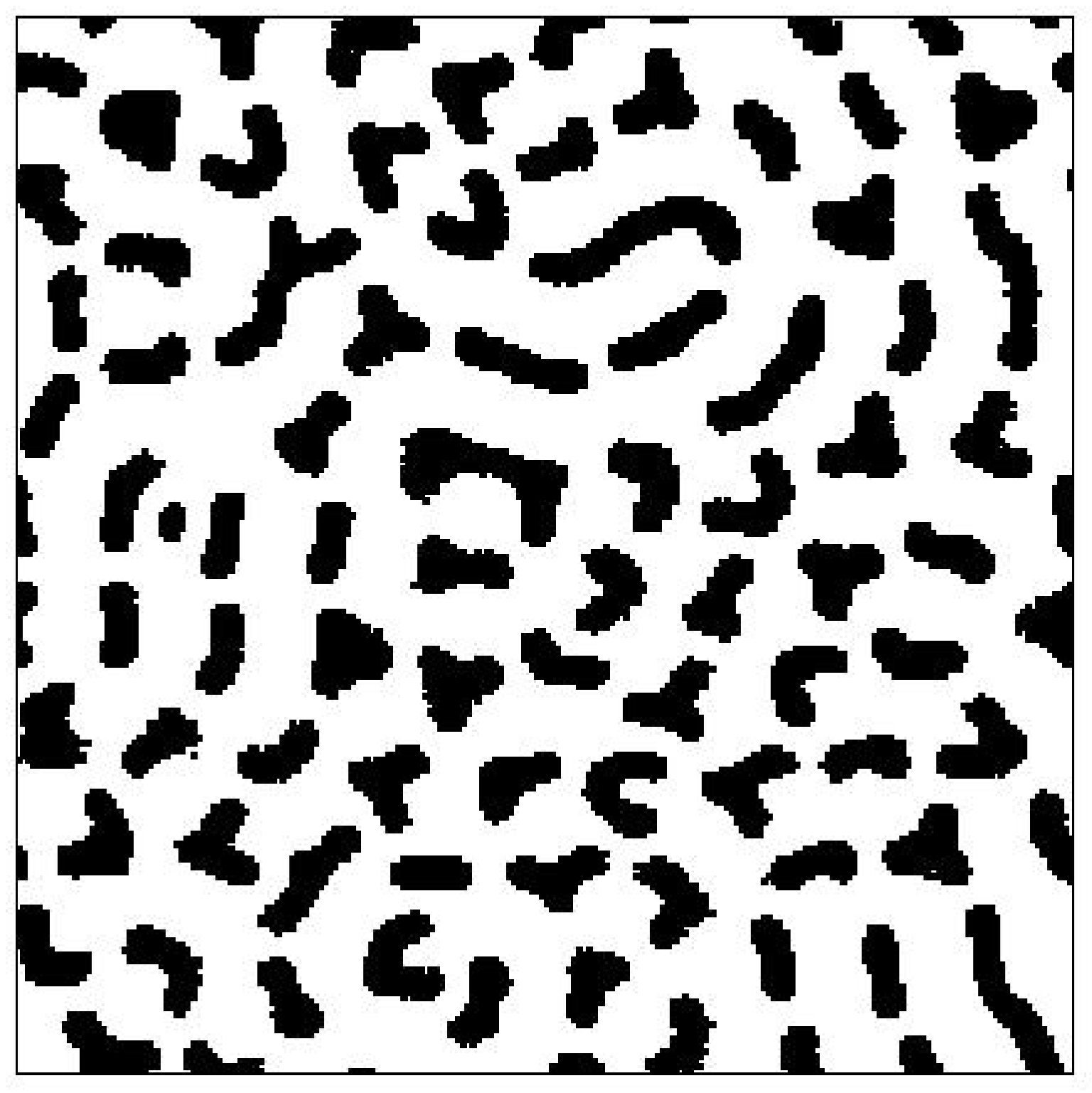} \\
\mbox{\bf (a)} & \mbox{\bf (b)}
\end{array}$
\end{center}
\begin{center}
$\begin{array}{c}\\
\includegraphics[width=7.5cm,keepaspectratio]{fig7c.eps} \\
\mbox{\bf (c)}
\end{array}$
\end{center}
\caption{(color online). The ``polycrystal'' microstructures. (a) The target medium, in which the
volume fraction of the black phase is $0.273$. (b) The reconstructed medium using the Yeong-Torquato
procedure \cite{Ye98a}. (c) The two-point correlation functions of the target and reconstructed media.
The mean squared error $E$ defined by Eq.~(\ref{eq_err}) is on the order of $10^{-7}$.} \label{fig7}
\end{figure}

We first consider an idealized microstructure of a two-dimensional ``polycrystal''
shown in Fig.~\ref{fig7}a. The ``polycrystal'' contains nine square  ``grains'', each associated with a
particular orientation as determined by the direction of the black ``stripes''.
The two-point correlation function of the black phase is shown in Fig.~\ref{fig7}c. It can be seen
that the oscillations of $S_2$ are the clear manifestation of the strong spatial
correlations between the ``stripes''. The reconstructed structure is shown in Fig.~\ref{fig7}b,
whose $S_2$ closely matches the target one with a very small error, i.e., sum of squared difference between the two,
on the order of $10^{-7}$. Though several stripe-like substructures have been roughly reproduced
and a few weakly preferred orientations of the substructures can be identified, the reconstruction
is distinctly different from the target structure in that it contains
much more compact substructures and the grain boundaries are completely missing. The correlations due
to the alternating ``stripes'' in the target medium are mimicked by the correlations between the
compact substructures with inter-spaces that are approximately equal to those between the ``stripes''.

\begin{figure}[bthp]
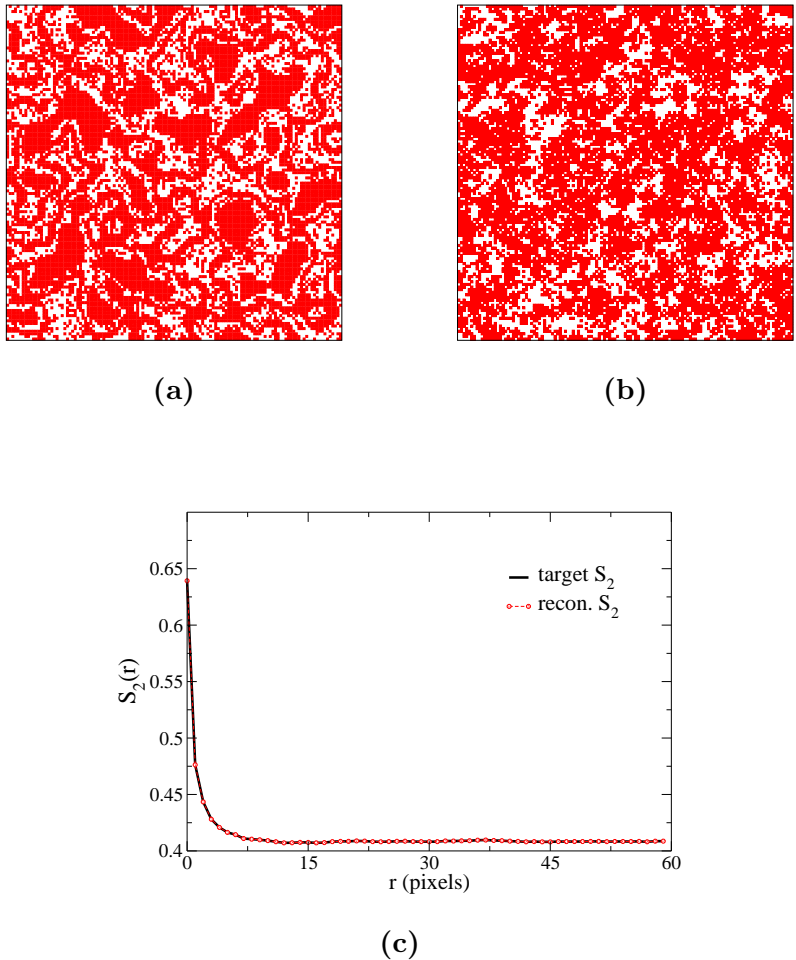

\begin{center}
$\begin{array}{c@{\hspace{1.5cm}}c}\\
\includegraphics[width=4.5cm,keepaspectratio]{fig8a.eps} &
\includegraphics[width=4.5cm,keepaspectratio]{fig8b.eps} \\
\mbox{\bf (a)} & \mbox{\bf (b)}
\end{array}$
\end{center}
\begin{center}
$\begin{array}{c}\\
\includegraphics[width=7.5cm,keepaspectratio]{fig8c.eps} \\
\mbox{\bf (c)}
\end{array}$
\end{center}
\caption{(color online). The laser-speckle patterns. (a) The target medium, in which the
volume fraction of the red phase is $0.639$. (b) The reconstructed medium using the Yeong-Torquato
procedure \cite{Ye98a}. (c) The two-point correlation functions for the red phase of the target and reconstructed media.
The mean squared error $E$ defined by Eq.~(\ref{eq_err}) is on the order of $10^{-7}$.} \label{fig8}
\end{figure}

Another interesting example is the reconstruction of a
laser-speckle pattern, which contains various shaped
sub-structures on a broad range of length scales \cite{Ji08} (see
Fig.~\ref{fig8}a). The two-point correlation function (shown in
Fig.~\ref{fig8}c) clearly does not reflect the multi-scale nature
of speckle pattern, which monotonically decreases to its long
range value very fast. The reconstructed pattern is shown in
Fig.~\ref{fig8}b, with a small error between the correlation
function on the order of $10^{-7}$. The substructures on different
length scales in the target medium (e.g., the disconnected compact
clusters, stripe-like structures and individual pixels) are all
mixed up in the reconstruction (e.g., the percolated clusters with
different sizes and shapes) to reproduce the short-ranged
correlations conveyed in $S_2$. The reconstructed pattern mimics
that of the Debye random medium, a famous model structure
containing ``clusters of all sizes and shapes'' with a exponential
two-point correlation function \cite{Ye98a, torquato}.

\begin{figure}[bthp]
\begin{center}
$\begin{array}{c@{\hspace{1.5cm}}c}\\
\includegraphics[width=4.5cm,keepaspectratio]{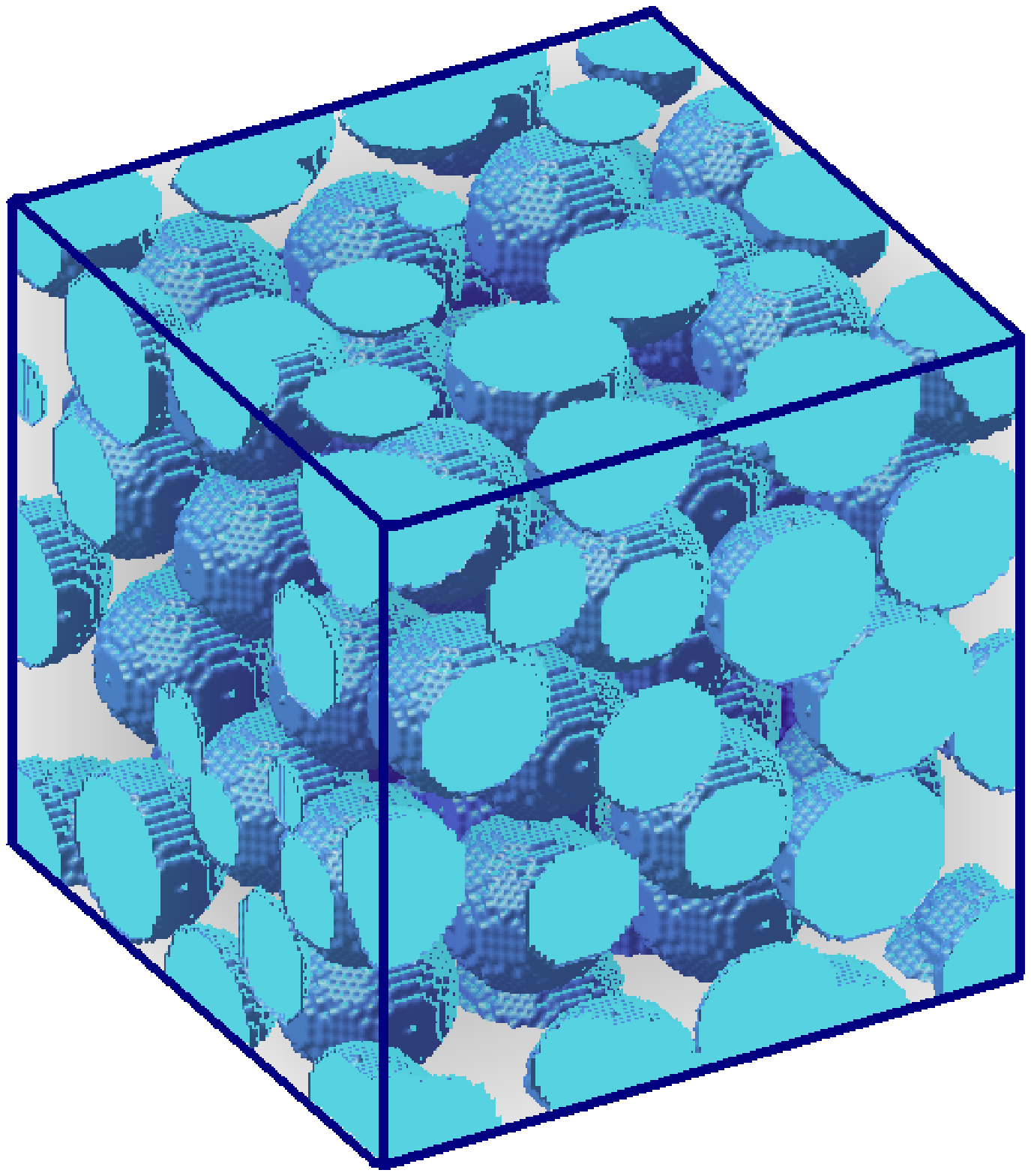} &
\includegraphics[width=4.5cm,keepaspectratio]{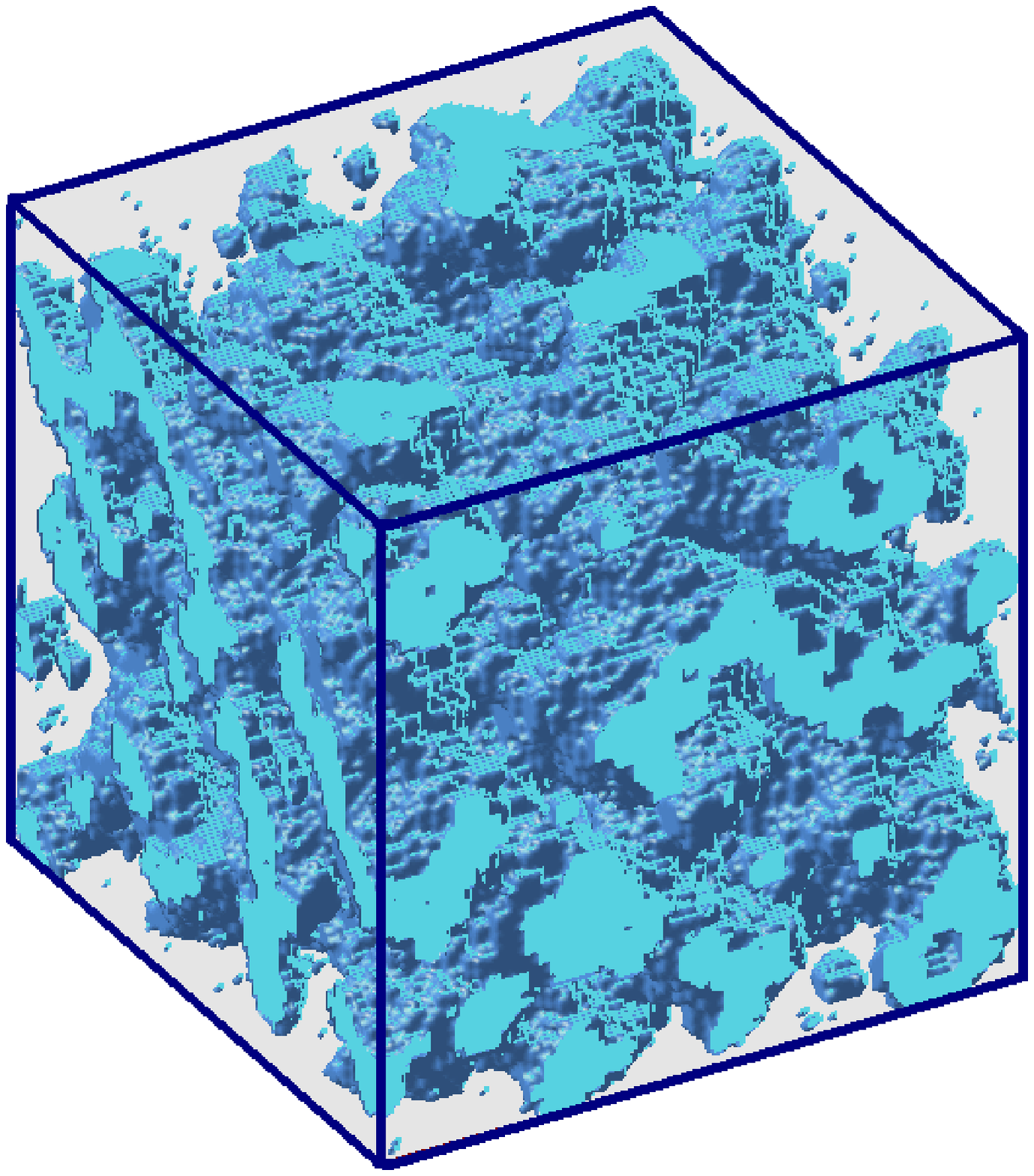} \\
\mbox{\bf (a)} & \mbox{\bf (b)}
\end{array}$
\end{center}
\begin{center}
$\begin{array}{c}\\
\includegraphics[width=7.5cm,keepaspectratio]{fig9c.eps} \\
\mbox{\bf (c)}
\end{array}$
\end{center}
\caption{(color online). The hard-sphere packings. (a) The target medium, in which the
volume fraction of the sphere (blue) phase is $0.446$. (b) The reconstructed medium using the Yeong-Torquato
procedure \cite{Ye98a}. (c) The two-point correlation functions of the target and reconstructed media.
The mean squared error $E$ defined by Eq.~(\ref{eq_err}) is on the order of $10^{-11}$.} \label{fig9}
\end{figure}

The above two example clearly show that $S_2(r)$ is not able to resolve the details of the substructures
in a complex medium. But even for simpler structures, $S_2(r)$ may still be insufficient.
Consider a realization of a three-dimensional equilibrium hard-sphere packing \cite{pnas} (see Fig.~\ref{fig9}a),
generated via standard Monte Carlo simulations. The volume fraction of the sphere phase
is close to the percolation threshold. The reconstruction is shown in Fig.~\ref{fig9}b, in which
the ``sphere'' phase forms a complex percolating structure. The error between the
target and reconstructed $S_2(r)$ is on the order of $10^{-11}$. It is clear that $S_2$
grossly overestimates the percolation of the target phases, which is due to its insensitivity
to the topological connectedness information of the media. This insufficiency of $S_2$ has
was pointed out previously \cite{Ye98a, Ye98b}.

Note that in the above examples, there are small but non-zero errors between
the target and reconstructed $S_2(r)$. 
This is an algorithmic implementation issue and can be removed if one used integers instead of
floating point numbers in the computer program. However, in real-world applications,
the obtained $S_2(r)$ data would in general suffer small but finite errors,
no matter how carefully the measuring experiments might be carried out. These examples clearly
reveal the insufficiency of $S_2$ in characterizing multi-scale structures and those
near percolation. Possible candidate correlation functions to overcome these shortcomings
are discussed in Sec.~IV.B. The examples given here are also approximate solutions of the invariance
equations (\ref{eq05}) derived in Sec.~II.B.

\section{Discussion}

\subsection{Generalization to Multiphase Media}

It is worth noting that although the focus of this paper is two-phase heterogeneous media,
it is straightforward to generalize the discussion to multiphase media. In
particular, consider a medium composed of $p$ distinct phases, each associated
with an indicator function ${\cal I}^{(i)}({\bf x})$ ($i=1, \ldots, p$) which equals
unity when ${\bf x}$ falls in phase $i$ and equals to zero otherwise. There are
in total $p^2$ two-point correlation functions $S^{(ij)}_2({\bf r})$ including
$p$ auto-correlations (i.e., when $i=j$ and the two points separated by displacement
${\bf r}$ falling into the same phase) and $(p^2 - p)$ cross-correlations (i.e.,
when $i \neq j$ and the two points falling into different phases $i$ and $j$).
However, there are only $p(p-1)/2$ independent correlation functions since the
$p$ indicator functions satisfy the equation

\begin{equation}
\label{eq401}
{\cal I}^{(1)}({\bf x})+{\cal I}^{(2)}({\bf x})+\cdots+{\cal I}^{(p)}({\bf x}) = 1,
\end{equation}

\noindent and thus, there are only $(p-1)$ independent ${\cal I}^{(i)}$.

For every one of the $p(p-1)/2$ independent two-point correlation functions, similar
feasibility and invariance conditions can be derived, which take the form of
integral or algebraic equations of the variations of the indicator functions
for the corresponding phases, as those derived in Sec.~II. We will not provide
the details of such derivations here. In addition, the same arguments concerning the
non-uniqueness of the solutions apply in the case of multiphase media, and thus
one cannot rule out the possibility of structural degeneracy.

\begin{figure}[bthp]
\begin{center}
$\begin{array}{c@{\hspace{1.5cm}}c}\\
\includegraphics[width=4.5cm,keepaspectratio]{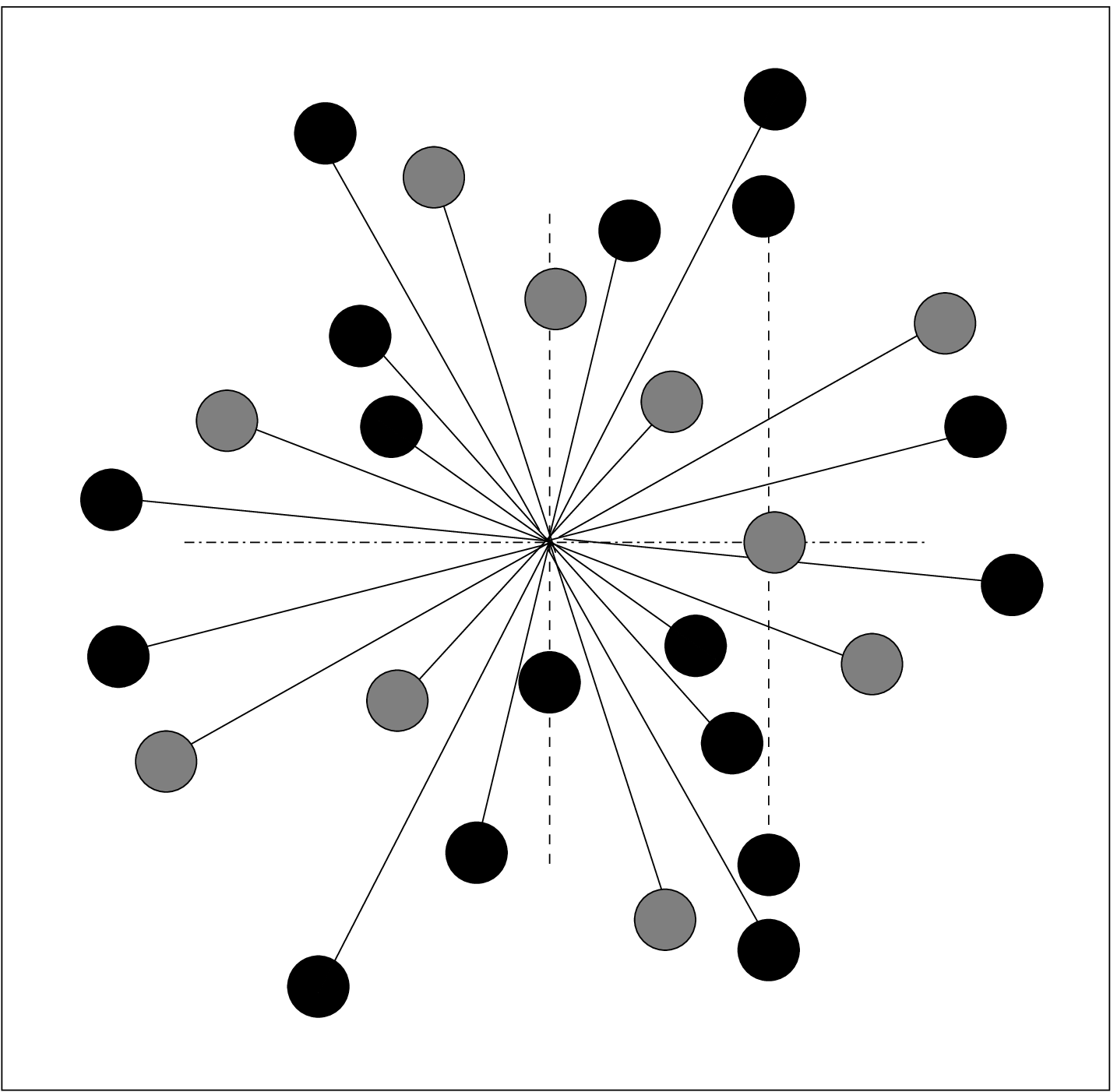} &
\includegraphics[width=4.5cm,keepaspectratio]{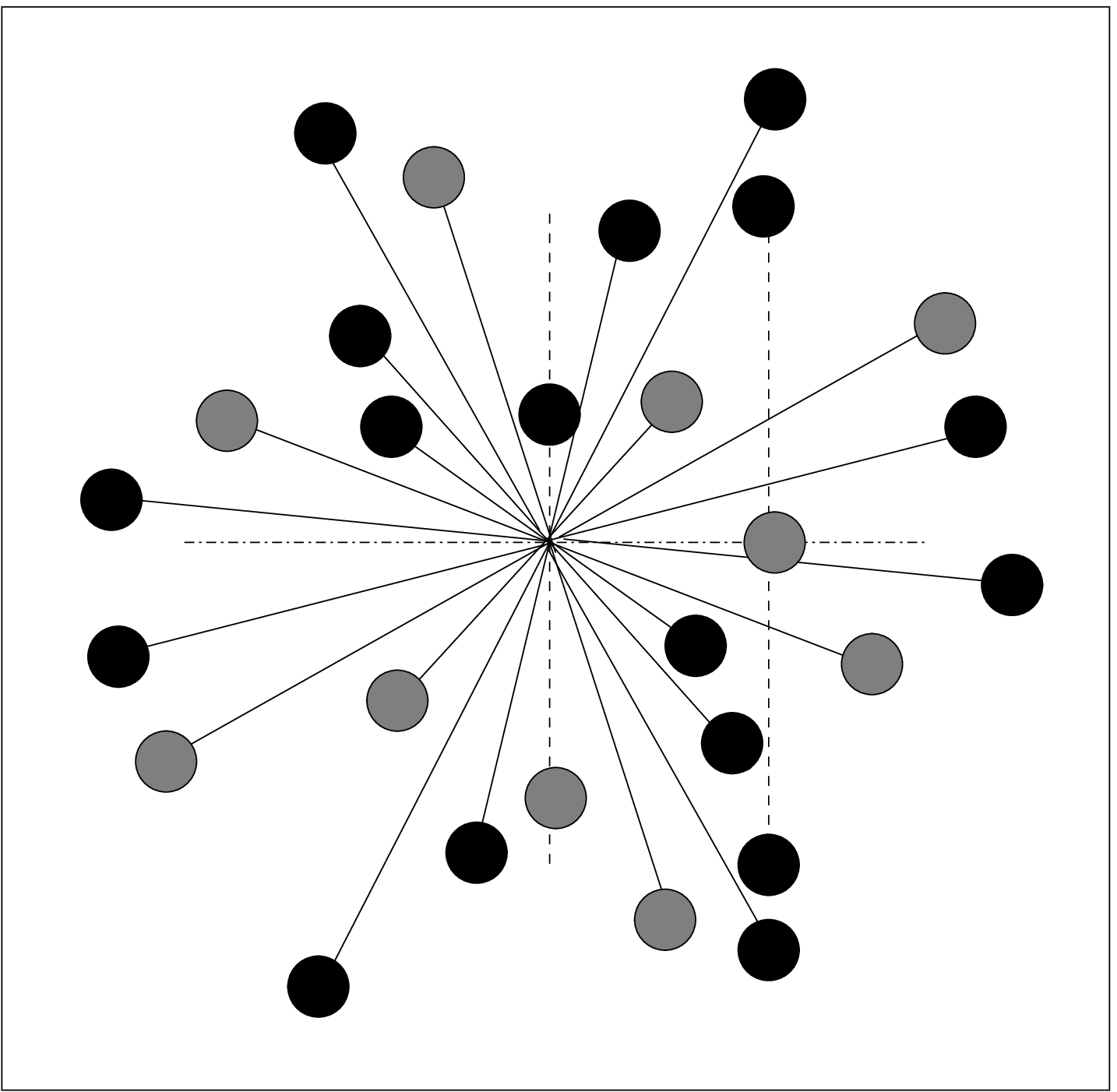} \\
\mbox{\bf (a)} & \mbox{\bf (b)}
\end{array}$
\end{center}
\caption{A degenerate pair of continuous three-phase media (black, white and gray) constructed based on
the degenerate circular-disk packings shown in Fig.~\ref{fig5}.} \label{fig10}
\end{figure}

\begin{figure}[bthp]
\begin{center}
$\begin{array}{c@{\hspace{1.5cm}}c}\\
\includegraphics[width=4.5cm,keepaspectratio]{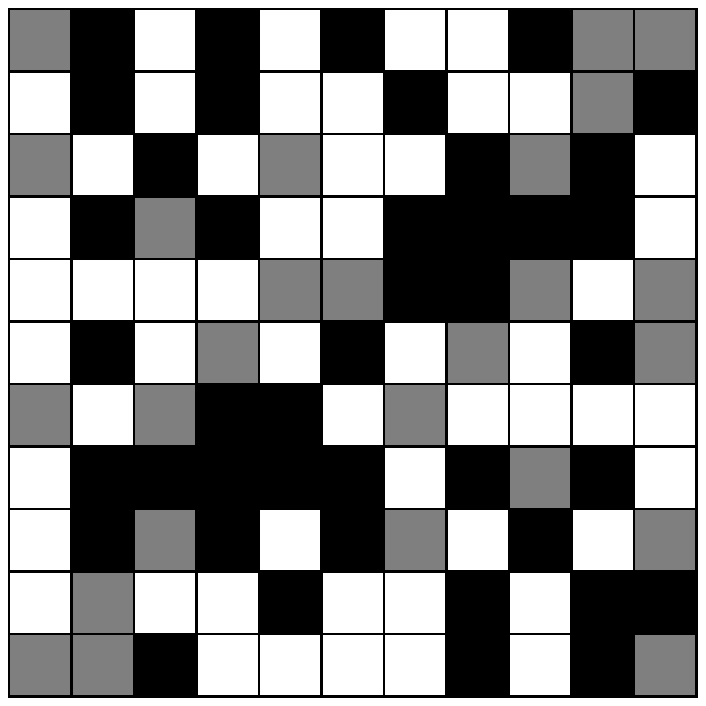} &
\includegraphics[width=4.5cm,keepaspectratio]{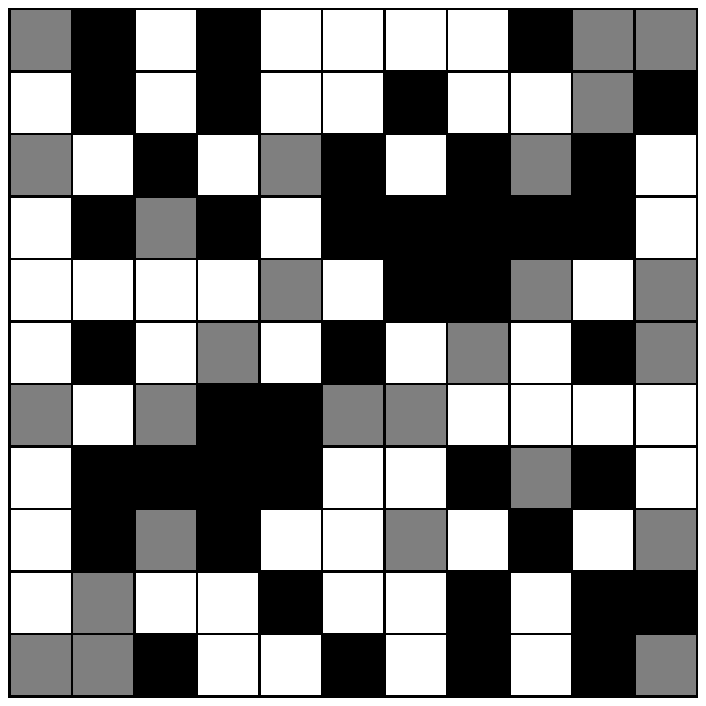} \\
\mbox{\bf (a)} & \mbox{\bf (b)}
\end{array}$
\end{center}
\caption{A degenerate pair of digitized three-phase media (black, white and gray) constructed based on
the degenerate media show in Fig.~\ref{fig6}.} \label{fig11}
\end{figure}

Two simple examples of degenerate three-phase media are shown in Figs.~\ref{fig10} and \ref{fig11}, which
 are constructed based on the circular-disk packings and their digitized analog given in
Sec.~IV.A, respectively. A subset of the particles (pixels) have been assigned to
the third phase, while all the pair-separation distances in the media remain the same.
Thus, any pair of independent two-point correlation functions of the two degenerate
media are identical.


\subsection{Additional Structural Information}


By examining the degeneracy conditions and constructing concrete examples,
we have established that two-point correlation function is in general
not sufficient to uniquely determine the structure of a heterogeneous medium and
degenerate structures do exist, especially for statistically homogeneous and
isotropic media. A natural question is what additional structural information
could be used to reduce the degeneracy.

\begin{figure}[bthp]
\begin{center}
$\begin{array}{c@{\hspace{0.75cm}}c@{\hspace{0.75cm}}c}\\
\includegraphics[width=4.25cm,keepaspectratio]{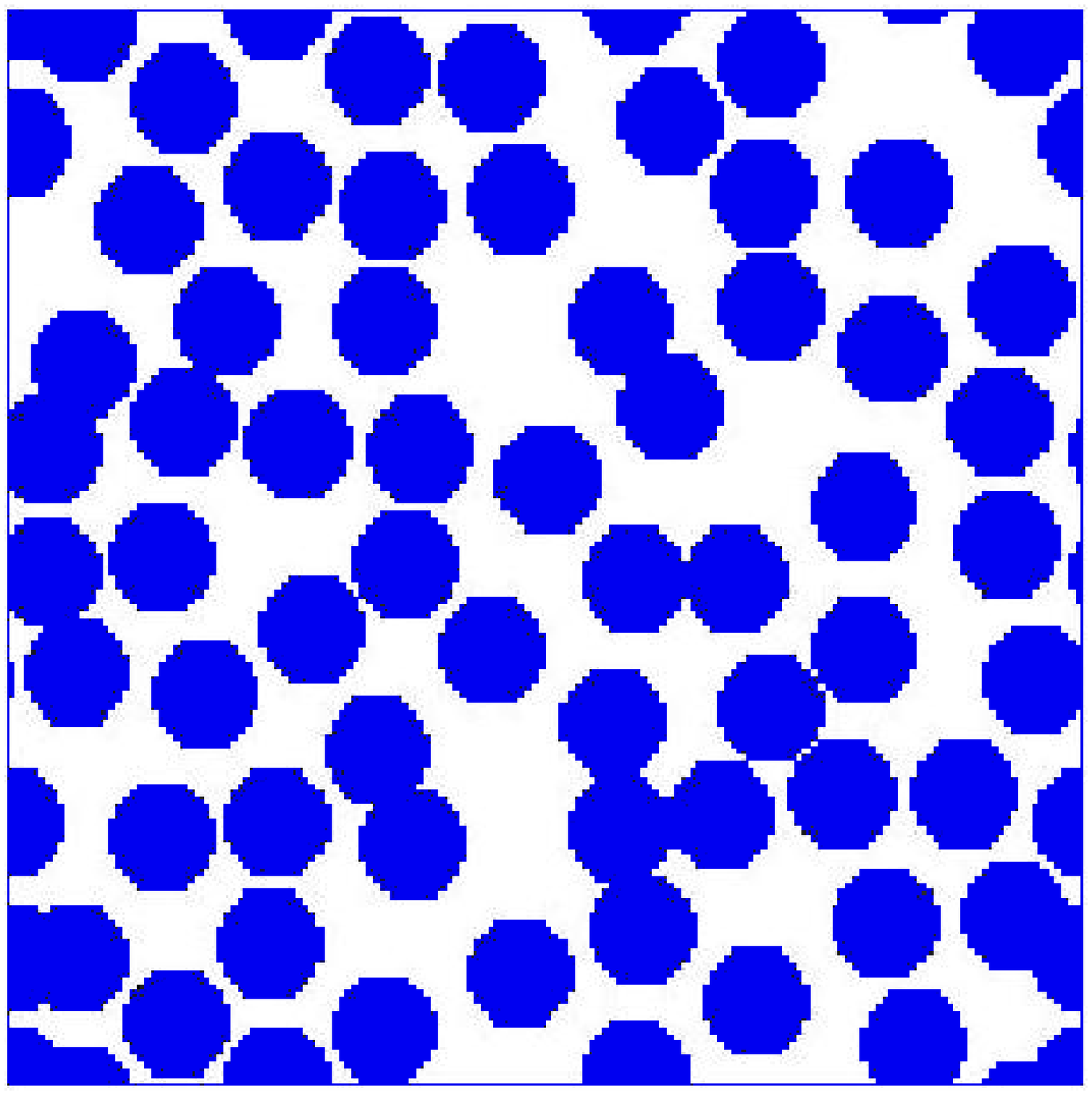} &
\includegraphics[width=4.25cm,keepaspectratio]{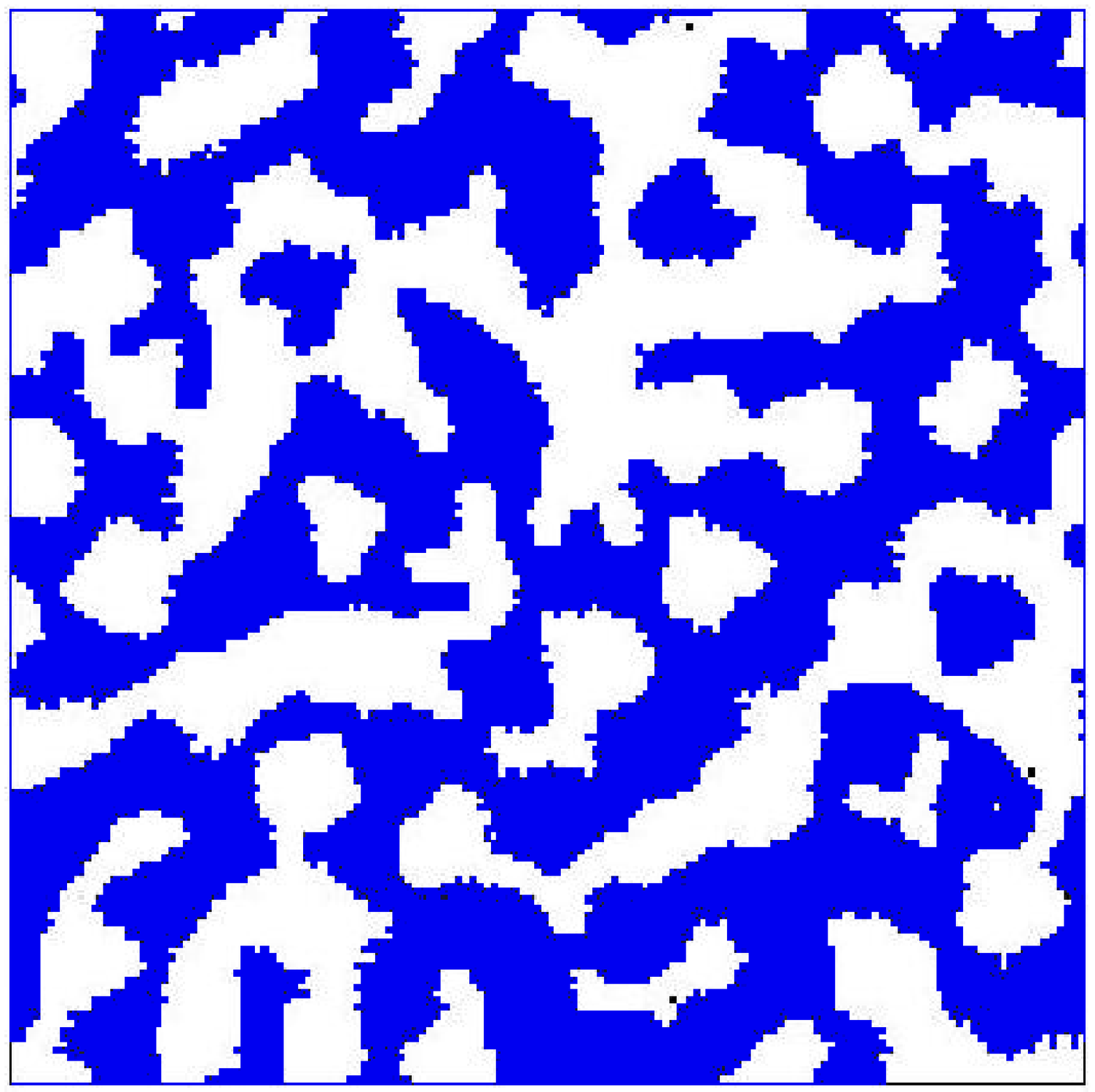} &
\includegraphics[width=4.25cm,keepaspectratio]{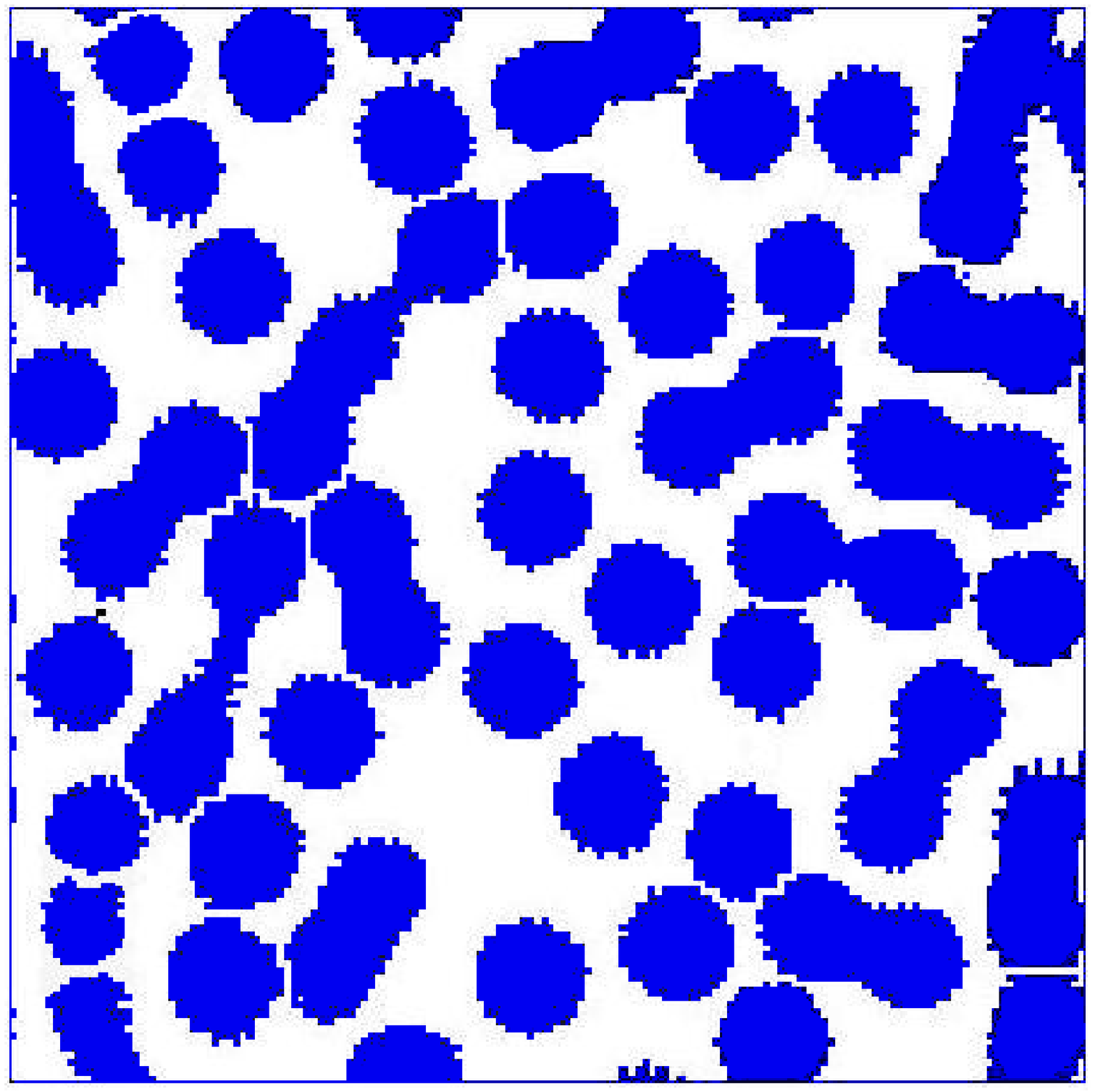} \\
\mbox{\bf (a)} & \mbox{\bf (b)}  & \mbox{\bf (c)}
\end{array}$
\end{center}
\caption{A circular-disk system containing small clusters. (a) Target medium, the volume
fraction of the ``disk'' (blue) phase is $0.532$. (b) Reconstruction using $S_2$ alone.
(c) Reconstruction incorporating both $S_2$ and $C_2$.} \label{fig12}
\end{figure}

It is notoriously difficult to find the complete answer to the above question.
However, we have shown in Ref.~\cite{pnas} that it would be a fruitful approach
to seek the most sensitive statistical descriptors among the various two-point
correlation functions (e.g., the surface functions and cluster-type functions \cite{torquato})
instead of using standard higher-order $n$-point correlation functions, as such $S_3$. The
two-point quantities are superior to $S_3$ in that their determination requires less
effort and they are sensitive to nontrivial structural information that
is not explicitly contained in $S_3$, such as the topologically connectedness information.

In particular, a superior descriptor proposed in Ref.~\cite{pnas} is the
two-point cluster function $C_2({\bf r})$, which gives the probability
of finding two points separated by ${\bf r}$ falling in the same cluster
of the phase of interest. A reconstruction example of a two-dimensional circular-disk
system containing small clusters from both $S_2$ and $C_2$ are shown in Fig.~\ref{fig12}.
It can be seen clearly that $S_2$-alone reconstruction grossly overestimates
the percolation of the ``disk'' phase, while the reconstruction incorporating
$C_2$ successfully reproduces the disks with small clusters.

We note that proper clustering information can also be used to
distinguish certain degeneracy examples provided in this paper.
Consider the degenerate media bearing a ``kite'' and ``trapezoid''
structure, respectively, shown in Fig.~\ref{fig1}, which are
obtained by decorating corresponding degenerate point
configurations. If we increase the size of the circular disks
while keeping their centers fixed, there exists a critical disk
diameter at which the three disks in the medium bearing the
``kite'' structure form a connected ``triangle'', which is
separated from the remaining disk; while in the medium bearing the
``trapezoid'' structure all the disks belong to a single linear
cluster. Because the connected ``triangle'' of disks only exists
in the medium bearing the ``kite'' structure, the degenerate pair
now can be distinguished, i.e., the degeneracy is reduced by using
the connectedness (i.e., clustering) information. It is clear that
in this case the clustering information reflects nontrivial
triangle information in the system, which is contained in $S_3$.

\section{Concluding Remarks}

In this paper, we discussed various aspects of the structural
degeneracy associated with the two-point correlation function
$S_2$ of heterogeneous media.
Complementary to previous studies, here we provide precise
mathematical formulations for structural degeneracy and rigorously show that
distinct media with exactly identical $S_2$ do exist.
In particular, we derived the exact
conditions for the existence of degeneracy in terms of integral
equations for continuous media and algebraic equations for their
digitized representations. By examining the equations and
constructing their solutions for specific examples, we have well
established that in general $S_2$ is not sufficient to uniquely
determine the structures of the media, contrary to previous claims
of the uniqueness of $S_2$-reconstructions based on numerical
studies. We have also provided a variety of concrete examples of
degenerate two-phase media, including both analytical
constructions and numerical simulations, which are respectively
exact and approximate solutions of the degeneracy equations. These
examples include analytically constructed patterns composed of
building blocks bearing the letter ``T'' and the word ``WATER'',
degenerate Barlow films as well as numerical reconstructions of
polycrystal microstructures, laser-speckle patterns and sphere
packings. It is clearly seen from these examples that $S_2$ alone
is unable to resolve the details of the microstructures and
usually overestimates the percolation in the media, which is
consistent with the results of our recent study on the
reconstruction of heterogeneous media using a wide spectrum of
statistical microstructure descriptors including $S_2$
\cite{pnas}. The conclusions also apply in the case of multiphase
media.

We have pointed out that it is necessary to include additional
information to better characterize the structure of
complex heterogeneous media beyond that contained in $S_2$. In Sec.~IV.B and Ref.~\cite{pnas}, we
have shown that the two-point cluster function $C_2$ (i.e.,
topologically connectedness information) is a superior
microstructure descriptor for media containing compact clusters
and its incorporation can significant reduce the degeneracy. For
more complex media such as those contain multi-scale
substructures, structural degeneracy associated with higher order
$S_n$ would exist, for example, we may find media that possess
distinct $S_4,~S_5,\ldots$ but identical $S_2$ and $S_3$. In
future work, we will focus on identifying such higher-order
degeneracies and seeking efficient statistical descriptors (e.g.,
higher-order versions of $C_2$) that can capture the salient
features of such media.

\begin{acknowledgments}
This work was supported by the
Office of Basic Energy Sciences, U.S. Department of Energy,
under Grant No. DE-FG02-04-ER46108.
\end{acknowledgments}


\begin{thebibliography}{10}




\bibitem{torquato}
S. Torquato, \textit{Random Heterogeneous Materials:
Microstructure and Macroscopic Properties} (Springer-Verlag, New
York, 2002).


\bibitem{Sa03}
M. Sahimi, {\it Heterogeneous Materials} (Springer-Verlag, New York, 2003).

\bibitem{Zo06}
T. I. Zohdi, Mech. Mater. {\bf 38}, 969 (2006).

\bibitem{Me07}
A. Mejdoubi and C. Brosseau, J. Appl. Phys. {\bf 101}, 084109 (2007).



\bibitem{sandstone}
D. A. Coker, S. Torquato and J. Dunsmuir, J. Geophys. Res. {\bf 101}, 17497 (1996).

\bibitem{Gi99}
L. J. Gibson and M. F. Ashby, {\it Cellular Solids}
(Cambridge University Press, Cambridge, England, 1999).


\bibitem{ecology}
A. Pommerening and D. Stoyan, Can. J. For. Res. {\bf 38}, 1110 (2008).

\bibitem{Pe93}
P. J. E. Peebles, {\it Principles of Physical Cosmology}
(Princeton University Press, Princeton, NJ, 1993).

\bibitem{Ga05}
A. Gabrielli, F. Sylos Labini, M. Joyce and P. Pietronero,
\textit{Statistical Physics for Cosmic Structures}
(Springer-Verlag, New York, 2005).

\bibitem{Kh08}
A. R. Kherlopian, T. Song, Q. Duan, M. A. Neimark, M. J. Po,
J. K. Gohagan and A. F. Laine, BMC Syst. Biol. {\bf 2}, 1 (2008).

\bibitem{cond}
M. Beran, Nuovo Cimento {\bf 38}, 771 (1965); S.
Torquato and J. D. Beasley, Inter. J. Eng. Sci. {\bf 24}, 415
(1986); S. Torquato and F. Lado, Proc. Royal Soc. Lond. A  {\bf 417}, 59 (1988);
D. C. Pham and S. Torquato, J. Appl. Phys. {\bf 94}, 6591
(2003).

\bibitem{elast}
M. J. Beran and J. Molyneux, Quart. Appl. Math. {\bf 24}, 107 (1966).
C. A. Miller and S. Torquato, J. Appl. Phys. {\bf 69}, 1948
(1991); J. Quintanilla and S. Torquato, J. Appl. Phys. {\bf 77},
4361 (1995); S. Torquato, Phys. Rev. Lett. {\bf 79}, 681 (1997).

\bibitem{fluid}
S. Prager, Phys. Fluids {\bf 4}, 1477 (1961).
J. D. Beasley and S. Torquato, Phys. Fluids A {\bf 1}, 199 (1989);
S. Torquato and B. Lu, Phys. Fluids A {\bf 2}, 487 (1990).

\bibitem{trap}
S. Torquato and J. Rubinstein, J. Chem. Phys.  {\bf 90}, 1644 (1989);
S. Torquato and F. Lado, J. Chem. Phys. {\bf 94}, 4453 (1991); S.
Torquato and D. C. Pham, Phys. Rev. Lett. {\bf 92}, 255505 (2004);
D. C. Pham and S. Torquato, J. Appl. Phys. {\bf 97}, 013535
(2005).

\bibitem{Re08}
M. C. Rechtsman and S. Torquato, J. Appl. Phys. {\bf 103}, 084901 (2008).

\bibitem{De49}
P. Debye and A. M. Bueche, J. Appl. Phys. {\bf 20}, 518 (1949).


\bibitem{To83}
S. Torquato and G. Stell, J. Chem. Phys. {\bf 78}, 3262 (1983).

\bibitem{To09a}
S. Torquato, Soft Matter {\bf 5}, 1157 (2009).

\bibitem{Ye98a}
C. L. Y. Yeong and S. Torquato S, Phys. Rev. E {\bf 57}, 495 (1998).

\bibitem{Ye98b}
C. L. Y. Yeong and S.  Torquato, Phys. Rev. E {\bf 58}, 224 (1998).


\bibitem{Cu99}
D. Cule and S. Torquato, J. Appl. Phys. {\bf 86}, 3428 (1999);
N. Sheehan and S. Torquato, J. Appl. Phys. {\bf 89}, 53 (2001).

\bibitem{Utz02}
M. G. Rozman and A. Utz, Phys. Rev. Lett. {\bf 89}, 135501 (2002);
D. T. Fullwood, S. R. Niezgoda, B. L. Adams and S. R. Kalidindi,
Prog. Mater. Sci. {\bf 55}, 477 (2010).

\bibitem{ApplyA}
K. Wu, M. I. J. Dijke, G. D. Couples, Z. Jiang, J. Ma, K. S. Sorbie, J. Crawford, I. Young and X. Zhang, Trans. Porous Media
{\bf 65}, 443 (2006).

\bibitem{ApplyB} M.~A.~Ansari and F.~Stepanek,
AlChE Journal, {\bf 52}, 3762 (2006).

\bibitem{ApplyC}
{R.}~{Hilfer} {and} {C.}~{Manwart},
{Phys. Rev. E} {\bf 64}, {021304} (2001).

\bibitem{ApplyD}
{D.}~{Basanta}, {M.~A.} {Miodownik}, {E.~A.} {Holm}, {and}
{P.~J.}~{Bentley}, {Metall. Mater. Trans. A} {\bf 36}, {1643} (2005).




\bibitem{ApplyE}
H. Kumar, C. L. Briant and W. A. Curtin, Mech. Mater. {\bf 38}, 818 (2006).


\bibitem{Ji07}
Y. Jiao, F. H. Stillinger and S. Torquato, Phys. Rev. E {\bf 76}, 031110 (2007).

\bibitem{Ji08}
Y. Jiao, F. H. Stillinger and S. Torquato, Phys. Rev. E {\bf 77}, 031135 (2008).

\bibitem{Ka08}
D. T. Fullwood, S. R. Niezgoda and S. R. Kalidindi, Acta. Mater. {\bf 56}, 942 (2008).

\bibitem{Ba04}
M. Boutin and G. Kemper, Adv. Appl. Math. {\bf 32}, 709 (2004).

\bibitem{Yel93}
J. I. Yellott, J. Opt. Soc. Am. A {\bf 10}, 777 (1993).

\bibitem{pnas}
Y. Jiao, F. H. Stillinger and S. Torquato, Proc. Nat. Acad. Sci. {\bf 106}, 17634 (2009) .


\bibitem{Ji09a}
Y. Jiao, F. H. Stillinger and S. Torquato, Phys. Rev. E {\bf 81}, 011105 (2010).


\bibitem{surface}
S. Torquato, J. Chem. Phys. {\bf 85}, 4622 (1986);
N. A. Seaton and E. D. Glandt, J. Chem. Phys. {\bf 85}, 5262 (1986);
C. Vega, R. D. Kaminsky and P. A. Monson, J. Chem. Phys. {\bf 99}, 3003 (1993).

\bibitem{lineal-path}
B. Lu and S. Torquato, Phys. Rev. A {\bf 45}, 922 (1992);
B. Lu and S. Torquato, Phys. Rev. A {\bf 45}, 7292 (1992);
D. Gueron and A Mazzolo, Phys. Rev. E {\bf 68}, 066117 (2003);
A Mazzolo, J. Phys. A {\bf 37}, 7095 (2004).


\bibitem{To93}
S. Torquato and B. Lu, Phys. Rev. E  {\bf 47}, 2950 (1993).

\bibitem{cluster}
S. Torquato, J. D. Beasley and Y. C. Chiew, J. Chem. Phys. {\bf 88}, 6549 (1988);
S. B. Lee and S. Torquato, J. Chem. Phys.  {\bf 91}, 1173 (1989).


\bibitem{accuracy}
It is only for certain simple idealized textures that one can
obtain a perfect match of the vector-argumented correlation
functions associated with the reconstructed and target media
\cite{Utz02}.

\bibitem{To99}
S. Torquato, J. Chem. Phys. {\bf 111}, 8832 (1999);
S. Torquato, Ind. Eng. Chem. Res. {\bf 45}, 6923 (2006);
J. A. Quintanilla, Proc. R. Soc. A {\bf 464}, 1761 (2008).

\bibitem{Yel92}
J. I. Yellott and G. J. Iverson, J. Opt. Soc. Am. A {\bf 9}, 388 (1992).

\bibitem{Ch00}
C. Chubb and J. I. Yellott, Vision Research {\bf 40}, 485 (2000).

\bibitem{ToJam07}
S. Torquato and F. H. Stillinger, J. Appl. Phys. {\bf 102}, 093511
(2007); S. Torquato and F. H. Stillinger, J. Appl. Phys. {\bf
103}, 129902 (2008).




\end{thebibliography}
\end{document}